\documentclass[fleqn,usenatbib]{mnras}

\usepackage[T1]{fontenc}

\DeclareRobustCommand{\VAN}[3]{#2}
\let\VANthebibliography\thebibliography
\def\thebibliography{\DeclareRobustCommand{\VAN}[3]{##3}\VANthebibliography}

\usepackage{graphicx}
\usepackage{amsmath}
\usepackage{amssymb}
\usepackage{multirow}
\usepackage{newtxtext,newtxmath}

\newcommand{\bpf}{{\texttt{BpF}}}
\newcommand{\rsp}{{\texttt{RSP}}}
\newcommand{\hecto}{Hectoshelle@MMT}
\newcommand{\hydra}{Hydra@WIYN}
\newcommand{\vdiv}{\rm div}

\title{Calibration of the convective parameters in  stellar pulsation hydrocodes}

\author[Kovács, Nuspl \& Szabó]{
Gábor B. Kovács,$^{1,2,3}$\thanks{E-mail: g.kovacs@astro.elte.hu}
János Nuspl,$^{2,3}$
Róbert Szabó$^{2,3,4}$
\\
$^{1}$ELTE Eötvös Loránd University, Institute of Geography and Earth Sciences, Department of Astronomy  1117, P\'azm\'any P\'eter s\'et\'any 1/A, Budapest, Hungary \\
$^{2}$Konkoly Observatory, Research Centre for Astronomy and Earth Sciences, E\"otv\"os Lor\'and Research Network (ELKH), MTA Centre of Excellence\\ H-1121 Budapest, Konkoly Thege Mikl\'os \'ut 15-17, Hungary\\
$^{3}$MTA CSFK Lend\"ulet Near-Field Cosmology Research Group,  H-1121 Budapest, Konkoly Thege Mikl\'os \'ut 15-17, Hungary\\
$^{4}$ELTE Eötvös Loránd University, Institute of Physics, 1117, P\'azm\'any P\'eter s\'et\'any 1/A, Budapest, Hungary
}

\date{Accepted 2023 March 17. Received 2023 March 16; in original form 2022 December 19}

\pubyear{2022}

\begin{document}
\label{firstpage}
\pagerange{\pageref{firstpage}--\pageref{lastpage}}
\maketitle

\begin{abstract}

Despite the appearance of two- and three-dimensional models thanks to the rapid growth of computing performance, numerical hydrocodes used to model radial stellar pulsations still apply a one-dimensional stellar envelope model without any realistic atmosphere, in which a significant improvement was the inclusion of turbulent convection. However, turbulent convection is an inherently multi-dimensional physical process in the vicinity of the ionization zones that generate pulsation. The description of these processes in one dimension can only be approximated based on simplified theoretical considerations involving several undetermined dimensionless parameters.

In this work, we confront two one-dimensional numerical codes, namely the  Budapest-Florida code (BpF) and the MESA Radial Stellar Pulsations module (RSP), with radial velocity  observations of several non-modulated RRab stars of the M3 globular cluster and specified the undetermined convective parameters by the measured data for both codes independently. 

Our determination shows that some of the parameters depend on the effective temperature, which dependence is established for the first time in this work, and we also found some degeneracy between the parameters. This procedure gives as by-product suggestions for parameters of the publicly available RSP code extensively used recently  by researchers through the MESA package.

This work is part of the preparatory work to establish a theoretical framework required to make progress based on the results of one-dimensional models to couple them with multi-dimensional ones for further detailed analysis of physical processes. 

\end{abstract}


\section{Introduction}

The theoretical modeling of classical radially pulsating stars has laid the foundations for the seismological study of stars and a better understanding of their internal structure. Over the last six decades, specific programs for numerical modeling  
been developed in different institutes and applied to interpret different measured data sets. These programs have been successful, but their application requires several specific knowledge and considerations and has been applied by small groups of researchers. 
In addition to the directly measurable quantities, e.g., pulsation period and amplitude, further individual considerations and additional independent codes are required for comparison with observations such as broadband photometric light curves, as the codes do not directly predict these quantities.
Their theoretical background came from diverse sources based on different and often incompatible assumptions.

The current abundance of mass survey data has created a need for codes based on well-established theoretical and numerical foundations which can be directly linked to measured data sets. This need calls for a review and improvement of the experience gained from previous developments to establish a solid basis for creating such a modern code and to make progress in multi-dimensional modeling and the description of recently observed but not yet modeled features.

The two main classes of radially pulsating stars, the Cepheids and RR Lyrae are different in mass and evolution state, but the mechanism causing and controlling their pulsation is almost identical. However, the RR Lyrae stars form a nearly homogeneous class in mass and luminosity with a narrow spread of low metallicity, so they served as examples at the dawn of modeling radial pulsation. They can also provide a base to extend the theoretical and computational background to improve modeling.

RR Lyrae stars arrive at the horizontal branch (HB) with similar parameters and structure. Their envelope  contains only a few percent of its mass but spans nearly half its radius. The pulsation happens in this region and is excited near the surface in the partial ionization zones. From a thermodynamic viewpoint, the system is open, transporting energy from the core to the surface; from a hydrodynamical point of view, it is in equilibrium in a stationary state. However, the temperature gradient is extremely steep in the ionization zones and unstable to convection both in equilibrium and in the instability strip evolving off the HB, where they show pulsation. We should solve the Navier-Stokes equations in three dimensions together with radiative transfer to follow the processes correctly in a stellar envelope. 

The compressible Navier-Stokes equations describe the movements down to the lowest spatial scale, incorporating convection and turbulence appearing. In practice, we have to apply approximate numerical methods to solve the equations to a reasonable resolution and a model for turbulence under this. In one-dimensional models, the 3-dimensional structure is mapped onto Lagrangian mass coordinates, and different models were constructed to describe the complicated turbulent convection in this framework. 

These one-dimensional models contain several undefined parameters up to 8, which cannot be deduced {\it ab initio} but only be determined through calibration to observations. Their application to different data sets showed that those could fit the data approximately with acceptable accuracy, but the best fits get to different parameters considering other data sets.
\footnote{For example, their dependence on effective temperature is shown later in this paper for RRab stars.}
Hence, these successes and the inherently multi-dimensional nature of the physical processes involved give us a ground base to extend the modeling into the multi-dimensional direction.

Toward this goal, we have to establish a theoretical framework to compare the results of one- and multi-dimensional numerical computations and the physical quantities not observable directly beside the physical processes governing them. In a series of papers, we will present our efforts toward this goal which is a theoretical investigation.

In this paper, we report on a first step of such a theoretical investigation considering a code we use (BpF) and the publicly available (RSP) code to scrutinize the physical quantities and processes in these models. As a first step, we had to specify the free parameters of the models with measurements to make the codes definite. With the least uncertainty, the radial velocity could be linked to the results of the model calculations, so we calibrated the models with these. 

First,  in Section \ref{sec:overview}. we summarize the successes of one-dimensional codes modeling radial pulsation, their main assumptions, and some of their deficiencies. Then in Section \ref{sec:model}. we deduce the common form of the equations used by the Budapest-Florida (hereafter \bpf ) and \texttt{MESA-}\rsp\ (in this paper: \rsp)\ codes. Based on these, we derived parameters from observations of several fundamental mode RR Lyrae stars of the M3 globular cluster. We describe the used data in Section \ref{sec:observations}. and present the fitting method in Section \ref{sec:method}. The results of the procedure are in Section \ref{sec:results}., which is discussed in Section \ref{sec:discussion}. In addition, we give a glimpse into the model  solutions through the variation of their internal structure to get a hint for possible progress to the solution to persisting problems, for which we show some examples in Sections \ref{sec:results-compare} and \ref{sec:discuss-diff} and backing its understanding with an attached animation of changes of the internal structure during pulsation.
 Finally we summarize our results and present a recommended convective parameter set for RRab stars in both codes in Section \ref{sec:summary}.

\section{An overview of one-dimensional radial stellar pulsation modeling}
\label{sec:overview}
Although the governing equations and numerical techniques were developed and updated during the last half century, the overall basic method of the modeling remained the same since \citet{Christy1964}.
\subsection{The basics of radial pulsation codes}
 The modeling consists of two main parts. 

The first step is to generate a starting envelope model, which can be used later for nonlinear calculations. Each hydrocode solves its governing equations in the hydrostatic form in this step, starting from the surface and proceeding inward. The surface condition is usually some top artificial mass element providing a constant pressure (i.e., ignoring any atmosphere model), while the inner boundary condition is a technically rigid region. The edge of this region is determined by the temperature of around $(2.5-5.0)10^6$ K. The reason for this approximation is that the pulsation does not penetrate the deeper layers of the star \citep{Catelan2015_book}. Thus the mass of the envelope is around 1\% of the star's total mass, while its spatial extent is around 50\% of the star's radius and well above its core. This fact means that effects like core-convection overshooting, nuclear processes, etc., have no direct influence on these envelope models on the timescale of the pulsation, and the inner region can be handled as a rigid sphere supplying a constant luminosity. 

From the point of view of pulsation, the convection and turbulence in the ionization zones are decisive due to the large temperature gradients. This region is not in thermodynamic equilibrium and is variable in its structure during the pulsation as the ionization fronts move out and backward, absorbing radiation at ionization and emitting radiation during the recombination. Hence their behavior is different from those present in the high-mass stars' cores or the low-mass stars' outer zones; also different from what might happen in the uppermost atmospheric layers. So, they determine, together with the opacity variations, the excitation and damping in a pulsating star with fine tuning of timing between them; this gives the 'time-dependent' tag. At the same time, we have to model the reduction of convection  into one dimension and the accompanying turbulence, which is a stochastic quasi-macroscopic problem, so the static models already contain the convection. 

In the second step, one calculates the linear nonadiabatic pulsation periods and eigenvectors and then creates a starting velocity profile from their linear combination. This way, the time to reach full amplitude pulsations is shortened as we do not have to wait for the damping of those overtones, which are damped anyway. Finally, we kick the static envelope with this velocity profile and start the time-dependent nonlinear calculation.

\subsection{From the codes to observables}
Another aspect of the modeling is the generation of synthetic light curves. In most cases, \cite{Stellingwerf1982b,bpf-beat2002}, the model light curve has larger amplitudes than the observed ones. The problem arises because the models are not calibrated to observations. After all, the pulsation codes usually give bolometric luminosities with diffusion approximation \citep{Stellingwerf1982a,Yecko1998,lengyel}. \citet{Vienna_Feuchtinger}  formulated a different solution (Vienna code), where this is not the case, and applies approximately calculated radiative transfer equations to generate light curves for RR Lyraes \citep{DorfiFeucthinger1999}, giving bolometric corrections for the Fourier-parameters of the light curves. \citet{Bono1994}  connected the slightly modified version of the \citet{Stellingwerf1982a} code to Kurucz static atmosphere models, under the assumption that while pulsation takes place in the envelope, its effects can be neglected in the outer atmosphere and it still can be approximated by quasi-static models providing synthetic colors.   A similar approach was used by \cite{Wood1997}.

\subsection{Development of codes}
The correct handling of  turbulent convection is one of the still open key questions in conventional numerical hydrodynamical codes used to model pulsating variable stars in the classical instability strip. 
The first models by \citet{Christy1964} completely ignored convective instabilities and used only one-dimensional (1D) radiative energy transport in the stellar envelope. His results described RR Lyrae stars well at the blue edge of the instability strip.  However,  the pulsation amplitudes in colder models were too high and showed instability beyond the observed edge. 

These deficiencies led to the efforts to include convective effects in the codes. The first attempt to perform this was using the static mixing-length theory \citep{mlt} by \citet{Baker-Kippenhahn1965}, which stabilized the red edge of the instability strip, but it was found at too low temperatures. The 'frozen-in' convection produced models that disagreed with observations \citep{Castor1971,Baker-Kippenhahn1965}, and it turned out that the inclusion of time-dependent convection is inevitable to get better models.
\footnote{ To keep the wording simple, in this paper'overshooting' means both 'turbulent convection in an extended region', and 'convective overshooting into a stable region.' Also see \citet{GW1992} and \citet{lengyel}.}

In the following two decades, several authors developed various treatments of 1D time-dependent convection \citep{Gough1977,Unno1967,Castor1968,Stellingwerf1982a}. For a good overview of these, see \citet{Baker1987}. \citet{Stellingwerf1982a} used a diffusion approximation for the turbulent energy and applied this approach in several models \citep{Stellingwerf1982b}. His work laid the foundation of later studies such as \citet{Kuhfuss1986} and \citet{Xiong1989}.
\citeauthor{Stellingwerf1982a}'s method was included in the adaptive grid used by \citet{GW1992} because the steep temperature gradient is only roughly resolved in the usually used 1D Lagrangian grids. Their theoretical analysis showed that in their model, it is possible to define the growth and decay rates of the turbulent energy, and the phase of the turbulent energy could precede and follow the opacity-controlled driving function. The overshooting behavior was also better described because, in the \citeauthor{Stellingwerf1982a} model,  turbulent energy was significant in the entire envelope of the star, so technically, the whole  envelope showed turbulent convection, not only the conventionally unstable region and its vicinity. Their method was adapted in the Budapest-Florida code by \citet{tcmodel}. 

The 1D treatment of turbulent convection brought in seven dimensionless free parameters describing physical processes, such as the timescale of the turbulent energy growth and decay, the mixing length, the constant eddy viscosity or the adiabatic compressibility index \citep{Yecko1998}. These authors tried to calibrate these parameters in their models while changing them, but a thorough calibration was neither published nor is it evident whether such a calibration is possible. Still, some frequently used combinations of the parameters, i.e. 'standard values', can be found in the literature. For example, these values differ slightly in RR Lyrae and Cepheid models to account for realistic radial velocity and luminosity amplitudes.

The Budapest-Florida code was the first successful attempt to model beat Cepheids \citep{bpf-beat2002} and double-mode RR~Lyrae stars \citep{bpf-drrlyr2004}. It turned out that beat Cepheid behavior comes naturally when turbulent convection is included \citep{bpf-beat2002} in the models. \citet{bpf-drrlyr2004} surveyed the RR Lyrae models for double-mode pulsations and investigated the topology of the RR~Lyrae instability strip from this point of view.

A very similar 1D pulsation code was developed by \citet{lengyel}, which was based on \citet{Kuhfuss1986} theory and was compared to the Budapest-Florida code. They used eight different free parameters similar to those of the \bpf 's \citep[for corresponding the two types of the parameters see][Table 1.]{lengyel}. These authors argued that the double-mode pulsation phenomena in the Budapest-Florida Code were artificial due to the incorrect treatment of the turbulent source function, which neglected negative (downward) buoyancy forces \citep{lengyel2}. They highlighted that this effect causes an artificial overshoot into deeper layers of the models, and that causes the double-mode pulsations; on the other hand, their models could not reproduce any double-mode pulsation.

\subsection{Multi-dimensional modeling}
One of the main reasons behind the 1D handling of the genuinely multidimensional convection phenomenon was the unavailability of the computational capacity needed for two- and three-dimensional calculations. The turbulent cascade goes through many orders of length scales, and its 3D hydrodynamic modeling in the modern era is still a challenging problem. Despite this, \citet{Deupree1977a,Deupree1977b,Deupree1977c,Deupree1977d} used a 2D hybrid (co-moving Lagrangian) approach to solve the Navier-Stokes equations directly. Although his work lacked good resolution, it showed that the convective cells could have the height of the convective zone \citep{Deupree1977d,Baker1987}. Recently \citet{SPHERLS1} developed a multidimensional pulsation code based on the original approach of \citet{Deupree1977a} called SPHERLS (SPHerically Eulerian Radially Lagrangian Scheme) for studying the interaction between convection and pulsation. This model was also used to fit a handful of M3 RR~Lyrae models \citep{SPHERLSIII,SPHERLS4} both in 2D and 3D, and the authors found that 3D calculations give almost the same result as the 2D counterparts with only a few percent difference in turbulent energy and amplitudes, for more detail see \cite{SPHERLS4}. 

There are other 3D convection models in the literature, although not on the topic of stellar pulsations \citep[for an overview in the topic of solar physics, see][]{Nordlund2009}. Most recently, \citet{Vasilyev2017,Vasilyev2018} used the Co5BOLD software \citep{Freytag2012} to model Cepheid spectra; meanwhile, \citet{Muthsam2010,kupka} developed a 3D hydrocode to study solar convection in a box on the surface of the Sun. Their code was adapted to Cepheid stars by \citet{Mundprecht2013,Mundprecht2015}, and used to calibrate the free parameter governing convective flux $\alpha_c$. They found that no single parameter would fit the 2D calculation and suggested that varying the dimensionless free parameters in time could be a better approximation in the 1D treatment.

As we can see, the multi-dimensional modeling of classical pulsators has been an emerging topic in the past few years. Yet, recently the use of 1D codes became easily accessible to the public by the inclusion of \citet{lengyel} code into the \texttt{MESA} \citep{Paxton2019} as the module Radial Stellar Pulsation (\rsp ). 

\subsection{Applications of the codes}

 However, these pulsation codes were not calibrated to observations, except for the code of \cite{Wood1997} and \cite{Bono1994}. The former was calibrated and used to model Bump 
Cepheid lightcurves \citep{KellerWood2002,KellerWood2006}. The latter was used extensively to model classical Cepheid light and radial velocity curves \citep{Marconi2013b,Marconi2013a,Marconi2017,Natale2008,Ragosta2019,DiCriscienzo2011}, as well as RR Lyrae stars \citep{Bono2000,DiFabrizio2002,Marconi2005,Marconi2007} and Bump Cepheids \citep{Bono2002}. 

Regarding the mixing length parameter $\bar{\alpha}_\Lambda=l/H_p$ \cite{Bono2002} found that they need to increase it from 1.5 to 1.8 to model Bump Cepheids closer to the red edge. \cite{DiCriscienzo2004,DiCriscienzo2004a} found that even larger ($\sim 2$) values may be justified for cooler RR Lyrae stars. Meanwhile, \cite{Marconi2013b,Marconi2017} found that in the case of some classical Cepheids, the projection factor (which term gives the connection between the measured radial velocities and spatial velocities) showed discrepancies when they tried to fit every photometric band and the radial velocities, usually closer the blue edge of the instability strip, which can be a sign that the underlying physical models of these codes may need corrections. \citep{bpf-beat2002} gave a trial-by-error based fixed parameter set for the convective parameters in the case of Cepheids and RR Lyrae stars, showing that there is probably no universal set for every type of pulsators. On the other hand, \cite{Paxton2019} only gives parameter prescriptions that are based on mixing-length theory, not observations. So in the case of the latter two codes, there is no general calibration by observations of the convective parameters, only some prescriptions, while the handling of negative buoyancy is also debated \citep[see e.g.][]{lengyel2}, and {\tt MESA-}\rsp\ is publicly accessible.

\section{The discussed two one-dimensional models}
\label{sec:model}
This section briefly compares \bpf\ and the  \rsp\  hydrodynamic codes. The models are based upon \citet{GW1992} and \citet{Kuhfuss1986}, respectively. \bpf\ uses a mix of treatments from \citet{GW1992} and \citet{Stellingwerf1982a}, which is described in detail in the original papers \citep[][]{bpf-beat2002}. The \rsp\ uses an almost identical treatment \citep{lengyel} with few differences based on \citet{Kuhfuss1986}. Here we reproduce the equations in a common form to clearly show the differences, and the details are given in the Appendix.

The equation of motion and conservation of energy are the following:
\begin{equation}
\label{eq:motion}
    \frac{du}{dt} = - \frac{1}{\rho} \frac{\partial}{\partial r}(p + p_t +p_\nu) - \frac{G M_r}{r^2} - \mathcal{D}_p
\end{equation}
\begin{equation}
\label{eq:energy}
\frac{de}{dt} + p\frac{dv}{dt} = - \frac{1}{\rho r^2}\frac{\partial}{\partial r} (r^2 (F_c + F_r - \mathcal{D}_F)) - \mathcal{C} 
\end{equation}
where $u$ is the velocity of the mass element, $\rho$ is the density, $r$ is the radius, $p$ denotes the pressure, $v$ is the specific volume, $e$ is the specific internal energy, $M_r$ is the stellar mass below the radius, $G$ the gravitational constant. The terms $\mathcal{D}_p$ and $\mathcal{D}_F$ are the different terms between the two hydrocodes. $F_c$ is the convective flux, while $F_r$ is the radiative flux, which is treated by diffusion approximation \citep{Yecko1998,lengyel}:

\begin{equation}
    F_r = 4r^2\pi\frac{4}{3} \frac{ac}{3\kappa}\frac{\partial T}{\partial M_r}
\end{equation}

The turbulent convection is included in these equations through the terms $p_t$, $p_\nu$, $F_c$, and with other terms as the turbulent energy $e_t$:

\begin{equation}
\label{eq:turbulent_energy}
    \frac{de_t}{dt} + (p_t +p_\nu)\frac{dv}{dt} = - \frac{1}{\rho r^2} \frac{\partial}{\partial r} (r^2 F_t) + \mathcal{C} + \mathcal{D}_e
\end{equation}

where $p_t$ is the turbulent pressure, $p_\nu$ is the eddy viscous pressure  (these two correspond to the Reynolds-tensors isotropic trace part and divergent part, for details see \cite{GW1992}),  $F_t$ is the turbulent flux. These convection quantities are expanded as follows:
\begin{align}
    &\label{eq:p_nu} p_\nu= - \frac{4}{3} \bar{\alpha}_\nu \rho \Lambda e_t^{1/2} r \frac{\partial}{\partial r}\left(\frac{u}{r}\right)\\
    &    p_t = \bar{\alpha}_p \rho e_t \label{eq:def-pt}\\
&    F_t = -\bar{\alpha}_t \rho \Lambda e_t^{1/2} \frac{\partial e_t}{\partial r} \label{eq:def-Ft}\\
 & F_c = \bar{\alpha}_\Lambda \bar{\alpha}_c\rho e_t^{1/2} \bar{Y} \label{eq:def-Fc}
\end{align}
where $\Lambda=\bar{\alpha}_\Lambda H_p$ is the mixing length, and  $H_p = pr^2/(\rho G M_r )$ is the pressure scale height.

The term $\mathcal{C}$ denotes the coupling between the turbulent and the internal energy and contains a turbulent source term $S$, a damping $D$, and a radiative damping term $\mathcal{D}_r$.
\begin{align}
\label{eq:coupling}
&    \mathcal{C} = S - D - \mathcal{D}_r \\
&    S = \bar{\alpha}_s\bar{\alpha}_\Lambda \frac{e_t^{1/2} p T Q}{H_p} \bar{Y} \label{eq:def-source}\\
&    D = \bar{\alpha}_d \frac{e_t^{3/2}}{\Lambda} \label{eq:def-damp}
\end{align}
where $Q=dv/dT$, and $\bar{Y}$ is the dimensionless entropy gradient. This latter term is the basic difference between the \bpf\  and \rsp.

The five remaining expressions ($\mathcal{D}_p$,$\mathcal{D}_F$,$\mathcal{D}_r$,$\mathcal{D}_e$,$\bar{Y}$) are different for the two models. We can separate the differences into three sets that are handled differently. The first group is a minor difference coming from the different handling of the deviating part of the Reynolds-tensor between \citet{Kuhfuss1986} and \citet{GW1992}. This has a small \citep[][section 5.2]{lengyel} difference in work and dynamics of the viscous stresses, denoted in our equations as $D_p$ and $D_e$:
\begin{align}
        &    \mathcal{D}_p = \left\{ \begin{array}{lr}
        0 &  \textrm{BpF}\\
        \frac{3p_\nu}{\rho r} &  \textrm{RSP}
        \end{array}\right. &         \mathcal{D}_e = \left\{ \begin{array}{lr}
        0 &  \textrm{BpF}\\
        \frac{3p_\nu u}{\rho r} &  \textrm{RSP}
        \end{array}\right.
\end{align}
The second difference group comes from the different radiative damping models, namely  the \bpf\ uses Péclet-correction \citep{bpf-beat2002,lengyel} and the \rsp\   \citep{lengyel} uses the method of \citet{Wuchterl1998}.
\begin{align}
\label{eq:dampr}
    &   \mathcal{D}_r = \left\{ \begin{array}{lr}
        \frac{D_r S}{1+D_r} &  \textrm{BpF}\\
        \frac{3}{4} \frac{e_t^{3/2}}{\Lambda}D_r &  \textrm{RSP}
    \end{array}\right. &     \mathcal{D}_F = \left\{ \begin{array}{lr}
        \frac{F_c D_r}{1+D_r} &  \textrm{BpF}\\
        0 &  \textrm{RSP}
        \end{array}\right.
\end{align}
Where $D_r$ is the reciprocal of the  Péclet number multiplied by the parameter $\bar{\alpha}_r$:
\begin{equation}
\label{eq:d_r_fac}
    D_r = \frac{4}{3}\bar{\alpha}_r \frac{acT^3}{\kappa \rho^2 c_p \Lambda e_t^{1/2}}
\end{equation}
where $c_p$ is the specific heat at constant pressure.

The last difference group is the definition of the dimensionless entropy gradient (or superadiabatic gradient) \citep{bpf-beat2002,lengyel}:
\begin{equation}
    \label{eq:Y_bar}
    \Bar{Y} =\left\{ \begin{array}{cc}
         \left[-\frac{H_p}{c_p}\frac{\partial s}{\partial r} \right]_+ &  \textrm{BpF}\\
         -\frac{H_p}{c_p}\frac{\partial s}{\partial r} &  \textrm{RSP}
    \end{array}\right.
\end{equation}

here $s$ denotes the specific entropy.

In the above equations, there are seven dimensionless parameters denoted by $\bar{\alpha}$-s (for the actual equivalences between the models and the common notation  as well as a short description, see Table~\ref{tab:unif_alpha}).
The parameters determine the efficiencies of the various physical terms in the description of the turbulent convection: the $\bar{\alpha}_\Lambda$ determines the mixing length of the convection, $\bar{\alpha}_c$ modifies the efficiency of the convective flux (see eq. \ref{eq:def-Fc}), $\bar{\alpha}_t$ gives the efficiency of the turbulent flux which term describes the diffusion of the turbulent energy \citep[see eq. \ref{eq:turbulent_energy}. and eq. \ref{eq:def-Ft}, theoretical explanation is given in][]{Kuhfuss1986}.The parameter $\bar{\alpha}_p$ is the scaling of the bulk turbulent stress (called turbulent pressure, eq. \ref{eq:def-pt}), while the parameter $\bar{\alpha}_\nu$ scales the eddy viscosity $p_\nu$ (the non-isotropic stresses caused by the turbulence, eq. \ref{eq:p_nu}). The remaining three parameters govern the coupling between the internal and turbulent energy: $\bar{\alpha}_s$ is the efficiency of the creation of convective eddies through buoyancy forces (i.e. the source of turbulence, eq. \ref{eq:def-source}), $\bar{\alpha}_d$ determines the efficiency of the dissipation of the turbulence (see eq. \ref{eq:def-damp}), and last but not least the $\bar{\alpha}_r$ parameter scales the radiative losses of the energy of the convective eddies.

These free parameters have the order of unity and are not calibrated by observations. \citet{bpf-beat2002} recommended two parameter sets for the \bpf\ code: one for RR~Lyrae stars and one for the Cepheids. \citet{Paxton2019} gives four parameter sets denoted by (A, B, C, D), but in the former case, the choice of the parameters was based upon approximate agreement with observed light curves, the sets in the latter case are more ad hoc. For example, the parameter set (A) corresponds to the standard \citet{Kuhfuss1986} mixing length, and multiple $\bar{\alpha}$-s are the same in those sets, having a preference towards the \citet{Kuhfuss1986} standards. More importantly, no investigation was carried out at that time regarding whether it is justified to use a given parameter set in a given pulsator type since there might be underlying correlations between physical quantities in the stars and the free parameters.  

\begin{table}

    \centering
    
    \caption{Unified $\alpha$ parameters}
    
    \label{tab:unif_alpha}
    \begin{tabular}{c|c|c|l}
        Unified & \bpf & \rsp &  Description \\
        \hline
        $\bar{\alpha}_\Lambda$ &  $\alpha_\Lambda$ & $\alpha$ & mixing length\\
        $\bar{\alpha}_\nu$ &  $\alpha_\nu$ & $\alpha_m$ &  scale of eddy viscosity\\
        $\bar{\alpha}_t$ &  $\alpha_t$ & $\alpha_t$ &  efficiency of turbulent flux\\
        $\bar{\alpha}_p$ &  $\alpha_p$ & $\alpha_p$ & scale of turbulent pressure\\
        $\bar{\alpha}_d$ &  $\alpha_d$ & $\alpha_d$ & efficiency dissipation of turbulence\\
        $\bar{\alpha}_s$ &  $\alpha_s^2\alpha_d$ & $\alpha_s$ & efficiency of source\\
        $\bar{\alpha}_c$ &  $\alpha_c$ & $\alpha_c$ & efficiency of convective energy transport\\
        $\bar{\alpha}_r$ &  $\alpha_r$ & $\gamma_r^2$ & efficiency of radiative losses\\
    \end{tabular}

\end{table}

In our comparison study, the models consist of 150 mass shells in both cases. In the starting model, the adaptive grid maintains the temperature at the 50th of mass shell counted from the surface to be $11000$ K. This shell is called the anchor zone \citep{Yecko1998,lengyel,Paxton2019}. This zone anchoring ensures that while iterating through the solutions of the static model, the spatial resolutions of the HI and HeI ionization zones remain unchanged. For more detail about these methods, we refer to the original papers of \citet{Yecko1998,bpf-beat2002,lengyel}.

\section{Observational data}
\label{sec:observations}

 Usually, for a comprehensive model calibration process, one uses multiple stars' simultaneous radial velocity and light curves to follow and reproduce phase lag and other effects and exclude the effects of changes in the pulsation. However, the radial velocity curve is better suited for fixating the models because it is measured and computed in absolute units (km/s) and comparable directly to model computations without any interpretation. At the same time, we have only bolometric luminosities in the hydrocodes without proper atmosphere models. 
 However, it is possible to model lightcurves \citep{Bono2000,KellerWood2002,Marconi2005,Marconi2007} or simultaneously RV curves too \citep{DiFabrizio2002,Natale2008,Marconi2013b,Marconi2013a,Marconi2017}, but this introduces additional assumptions (e.g. static atmosphere approximation), and it is done with neither of the studied pulsation codes in this paper.

Because of this, we use selected RR Lyrae stars of the globular cluster M3 measured by \citet{Jurcsik2017} using various types of equipment: \hydra\ \citep{Hydra2001} and \hecto\ \citep{Szentgyorgy2011} for the RV measurements and simultaneous BVI$_c$ photometric observations taken by the  60/90 Schmidt telescope at Konkoly Observatory \citep{Jurcsik2015}.

We have selected eight input stars from the observed sample covering the temperature range of $[6300;6800]$ K. The base of the selection criterion was that the stars had an excellent \hecto \ coverage for the entire radial velocity curve, especially in the short ascending phase, since \hecto\  data have a much smaller error than \hydra (1 km/s vs. 5-7 km/s). We excluded all Blazhko-stars because the hydro models used in this work cannot reproduce the Blazhko-effect.

\begin{figure}
    \centering
    \includegraphics[width=\columnwidth]{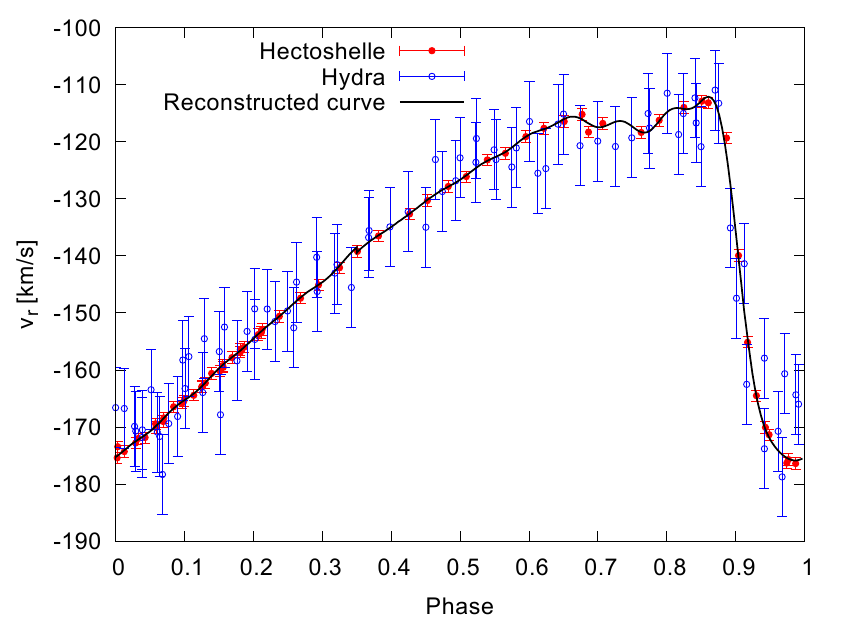}
    \caption{Phase folded raw radial velocity data and reconstructed curve from \citet{Jurcsik2017} in the case of v036, an RRab star in the globular cluster M3. Blue empty circles denote \hydra\ measurements, and red filled circles mean \hecto\ measurements. The black line is the whitened Fourier-reconstructed curve sampled equidistantly that we use for the calibrations after removing the mean and applying the p-factor.}
    \label{fig:reconstructed_curve}
\end{figure}

We use the radial velocity data in the form used for Baade-Wesselink method by \citet{Jurcsik2017}. This means that based on the \hecto\  measurements, we clean and re-sample the data by Fourier decomposition and keep the first 15 harmonics of the curve. One example of the resulting curves can be seen in Fig.~\ref{fig:reconstructed_curve} for the star v036. The uncertainties remain below 1 km/s,  and the equidistant sampling means there will be no need to weight the data points by phase coverage.

The input parameters of the models consist of the bolometric luminosity, effective temperature, mass,  hydrogen, and metal content. We use the temperatures as given in \citet{Jurcsik2017}, and to calculate their bolometric luminosity, we use the mean V magnitude with the distance data from \citet{Jurcsik2017} (Gaia EDR3 provides similar errors, and the mean distances are also consistent with that of \citet{Jurcsik2017}), the bolometric corrections of \citet{Torres2010} and interstellar extinction from \citet{Schlafly2011}. These initial parameters are shown alongside the selected stars in table \ref{tab:stars} , and the estimation method for the remaining three input parameters are described in Section~\ref{sec:method}. We note that our $L_{\rm bol}$ values are somewhat lower than those in the work of \citet{Cacciari2005}. These differences can result from the difference in the temperature and mean V magnitude data, as we have not included other instruments and data in our calibrations. We handle the effect of the uncertainties in the luminosities in section \ref{sec:results}.

The p-factor has an essential role in the RV curves.  There is an ongoing debate about the period dependence of the projection factor \citep[e.g. see][]{Molinaro2012,Marconi2013a} although the effect of this period dependence in the case of RRab stars is much smaller due to the relative similarities of the periods. In the case of classical and Bump Cepheids using p-factor as a free parameter in model fitting \citet{Marconi2013a,Marconi2017} found unrealistically low p-factors ($p\sim 1$) in some cases. Therefore we constrained it to a fixed value of
 $p=1.34\pm0.07$  by using the (p-P)-relation from \citet{Nardetto2009} substituting $P=0.5$, the characteristic period of RR Lyraes; this value is equal to $p=1.35$ within the error used by \citet{Jurcsik2017} based on \citet{Nardetto2004} , as well as to the value $p=1.36$ adopted for the CM Leonis by \citet{DiFabrizio2002}, and is the same as the empirical value from \cite{2018pas6.conf..213T}.

\begin{table}
    \centering
    \caption{The used stars from \citet{Jurcsik2017}, and the selected input model parameters.}
    \label{tab:stars}
    \begin{tabular}{c|ccccccl}
        Star & 	Period & $T_{\textrm{eff}}$ &	$L_{bol}$ & $M$ & $X$ & $Z$\\
          ID & 	[d] & [K] &	$[L_{\odot}]$ & [$M_\odot$] &  & \\        
        \hline
        v009&  	0.541549	&$6619$&	$47.5 \pm 2.8$ & 0.63631 & 0.76 & 0.0004\\
        v018& 	0.516455	&$6681$&	$41.0 \pm 2.5$ & 0.54180 & 0.75 & 0.0004\\
        v019& 0.631980&$6377$&    $48.3 \pm 2.8$ &0.62683 & $0.72$ & $0.0010$\\
        v036& 	0.545596&$6678$&	$44.2 \pm 2.7$ & 0.54796 & 0.77 & 0.0005\\
        v046& 	0.613388&$6394$&	$43.8 \pm 2.7$ & 0.57116 & $0.73$ & $0.0006$\\
        v060&  0.707729&$6461$&	$51.8 \pm 3.2$ & 0.53306& $0.75$ & $0.0006$\\
       
        v083& 0.501270& $6750$ & $45.7 \pm 2.7$ & 0.61705 & $0.75$ & $0.0006$\\
        v120& 0.640145&$6326$& $46.3 \pm  2.8$ & 0.60817 & $0.73$& $0.0009$ \\
    \end{tabular}

\end{table}

\section{Fitting method}
\label{sec:method}

We calibrate the convective $\bar{\alpha}$ parameters by running the two nonlinear numerical codes on multiple ($\bar{\alpha}$) parameter  grids using the observed parameters of the eight chosen stars and fit both resulting phase-folded  model RV curves to observations of \cite{Jurcsik2017}  based on the reduced $\chi^2$ values. The input parameters of the hydrocodes are the bolometric luminosity, stellar mass, hydrogen, helium, and metal contents. At the same time, the bolometric luminosity was acquired directly from the observations described in the previous section; the remaining parameters must be determined in a self-consistent way, as there are no exact values in the literature. 

The (X,Y) contents and metal abundance were not taken simply by choosing values in agreement with the [Fe/H] in the literature but searched for accurate values in an interval around it. The reason for this assumption is that the spectroscopic and photometric [Fe/H] values for M3 RR Lyrae stars show discrepancy \citep{Jurcsik2003}. In a recent study, two distinct metallicity groups at [Fe/H]$\approx -1.6$ and [Fe/H]$\approx -1.45$ were established by \citet{Lee2021}. 
Otherwise, the helium abundance strongly affects the amplitude and luminosity of RR Lyrae stars \citep{Marconi2018}. Since we have no available data about the helium abundance of these stars, we use a grid of (X,Z) pairs for $X \in \{0.70;0.80\}$ with $0.01$ steps and $Z \in \{0.0001;0.001\}$ with $0.0001$ steps, around [Fe/H]$=-1.5$. We discuss the possible introduced errors by this method in Section \ref{sec:errors} .

Secondly, the mass of the star has a strong influence on the pulsation period \citep{Marconi2015,Marconi2018} which can also be calculated for a star by linear-non-adiabatic (LNA) models\footnote{Strictly speaking, the linear and nonlinear periods are different, but this difference is very small (see Section~\ref{sec:errors}), while the use of LNA model calculation is straightforward since the LNA eigenvectors are used to start the non-linear calculation \citep{bpf-beat2002,lengyel}, and calculation time of an LNA model is magnitudes smaller  than a nonlinear run (seconds compared to hours).}.
Hence, to acquire mass values, we scanned an LNA model grid calculation for the periods in the range $M\in [0.4,0.8] M_\odot$ and each (X,Z) pair. Then we compared the measured periods with the period-mass data of this grid to obtain estimated mass values. Based on these values, we calculated full amplitude nonlinear models for each $(X,Z)$ pair and selected the one with the lowest $\chi^2$ calculated for the RV curve. The stellar parameters chosen in this way were the input parameters of the calibration process and are shown in Table~\ref{tab:stars}.  We note that our input parameters are consistent with those that can be derived from the PLMTZ and PLMTXZ relations of \cite{Marconi2015} and \cite{Marconi2018}.

For the calibration of the $\alpha$ parameters, we fit the calculated model radial velocity curve to the cleaned observed ones, excluding the minimum velocity phase and bump, as velocity values are more dependent on the used metallic lines and the actual method of the radial velocity determination (see \citealt{Anderson2018} for an overview). 
To fit the ascending and descending phases of the RV curve and the velocity maxima, however, one needs to consider the overall collection method of RV curves.
The radial velocity measurements are usually collected by averaging the individual RV data 
of multiple absorption lines. These lines are usually formed at slightly different optical depths, meaning that the measured RV curve is not connected to a single optical depth but to an average \citep{Anderson2018}. To account for this, we sample the model's RV profile at different optical depths with the same distribution and weights as in the original paper of \citet{Jurcsik2017}. We also find that if one does not use weights, a simple mean of the RVs at optical depths ranging from $0.01$ to $2/3$, the resulting RV curve won't deviate from the one calculated with weighting by more than 5\%. 

The convective parameters have different effects on the radial velocity and light curves. Although we choose to calibrate our $\alpha$ parameters with the help of the radial velocity curves, here we demonstrate their effects on the light curves for  qualitative comparison. To produce synthetic light curves, we use the linear bolometric corrections for the Fourier parameters of the first five overtones described by \citet{DorfiFeucthinger1999} applied on the photospheric model light curve calculated at $\tau=2/3$ optical depth. 
In our study, we experienced that their corrections can be used in most cases. The only exception is the low  eddy viscosity (i.e., $\bar{\alpha}_\nu < 0.1$) regime where the light curves require more than 15 harmonics to be described adequately.

\begin{figure*}
    \centering
    \includegraphics[width=\textwidth]{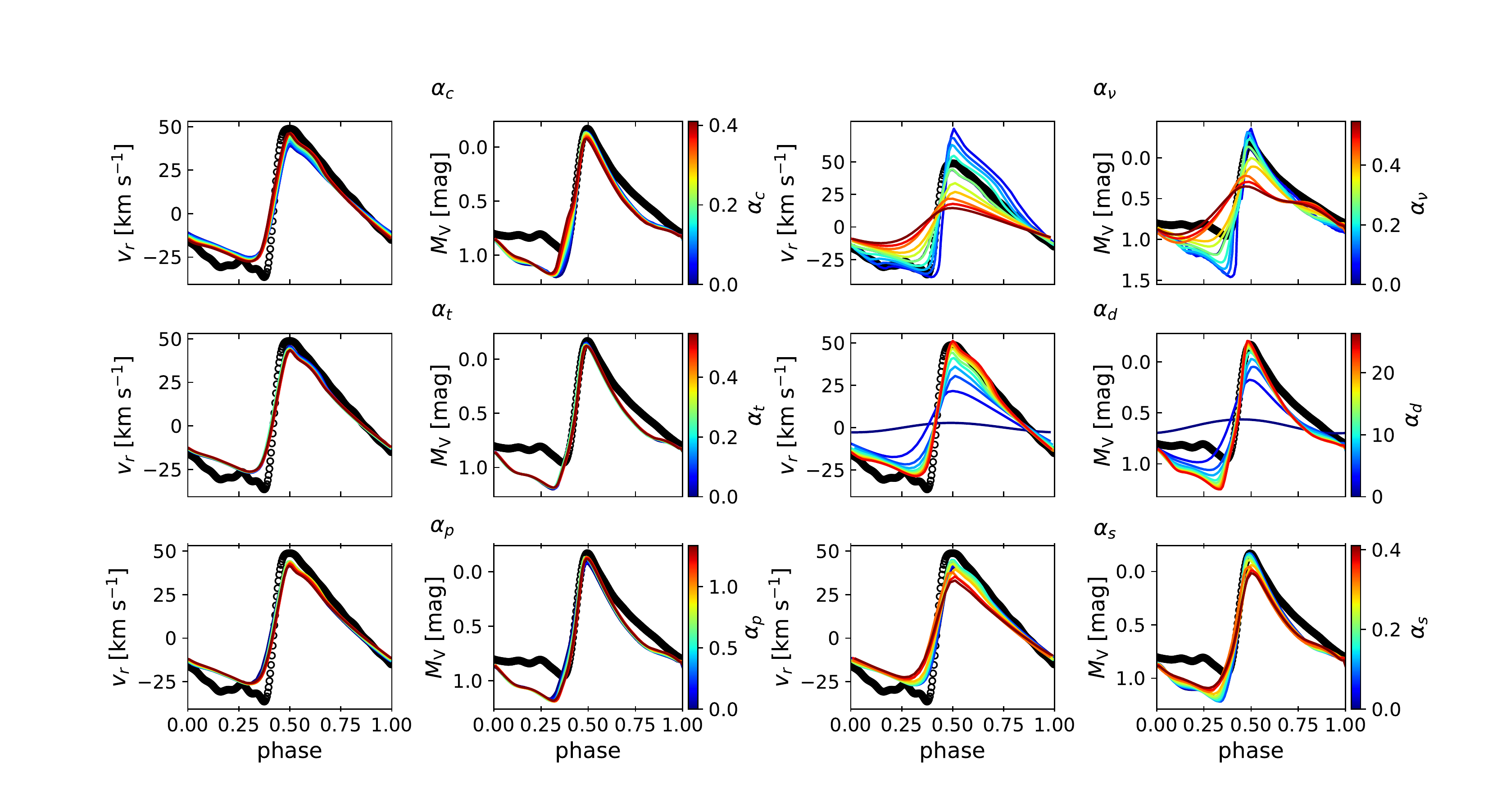}
    \caption{Effects of the convective parameters on the radial velocity and light curves in the case of the star v036 with the \bpf\ model. In every horizontal pair of panels, the left-hand side shows the radial velocity curve versus the phase, and the right-hand side panel shows the phase folded absolute V light curve. Every model is calculated by changing only one parameter while keeping the others at their prescription values \citep{bpf-beat2002}. Black circles are the observed values. Color denotes the value of the changed parameter as shown by the color bars. The changed parameters are (from top to bottom, left to right): $\alpha_c$,$\alpha_p$,$\alpha_t$,$\alpha_\nu$,$\alpha_d$,$\alpha_s$.}
    \label{fig:bpf_parameter_change}
\end{figure*}
\begin{figure*}
    \centering
    \includegraphics[width=\textwidth]{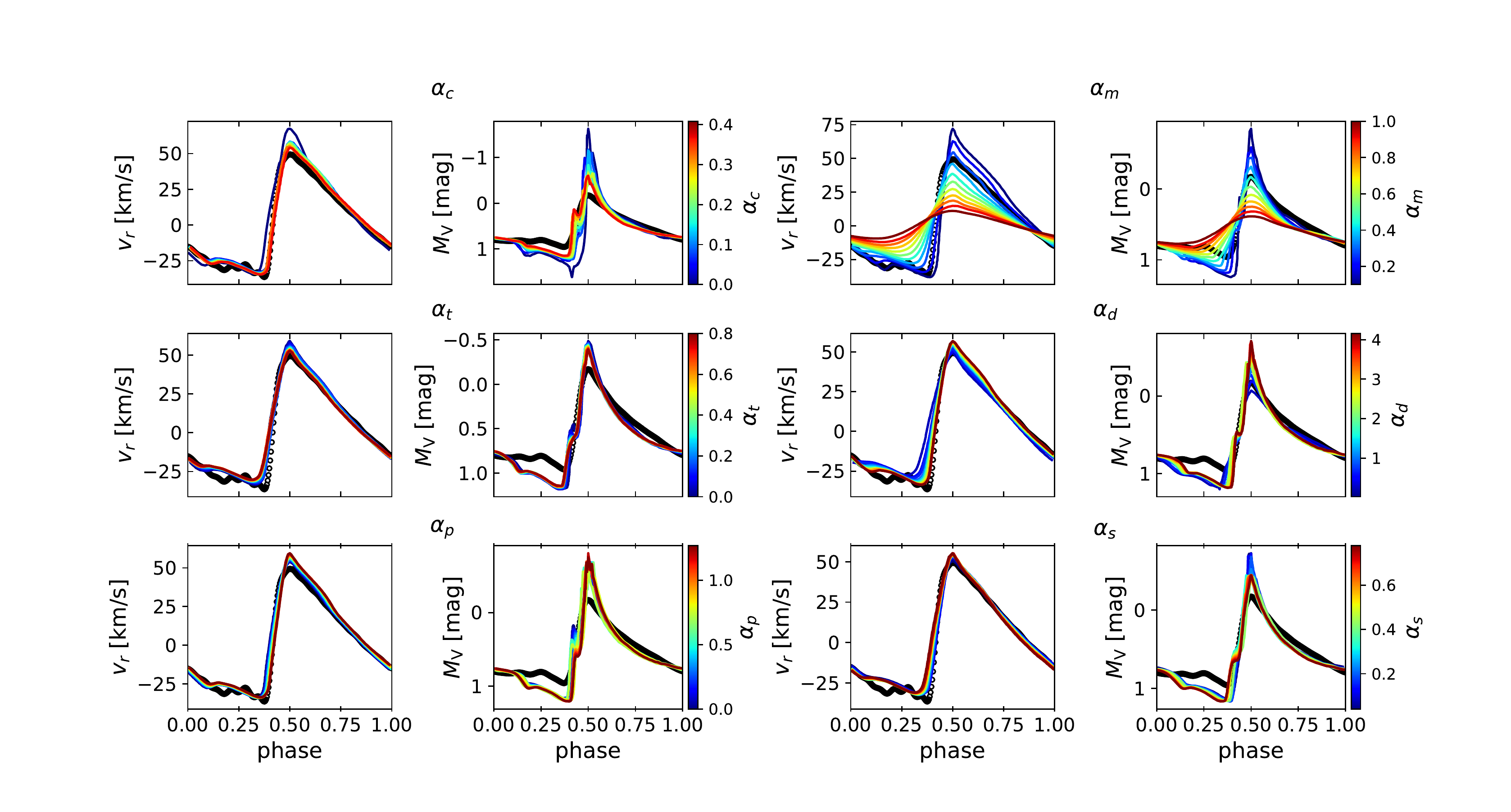}
    \caption{The same as Fig.~\ref{fig:bpf_parameter_change}. but in the case of the \rsp\ models. The prescripted values come from RSP parameter set C \citep{Paxton2019}.}
    \label{fig:rsp_parameter_change}
\end{figure*}

The greatest obstacle to our parameter calibration is the sheer number of the parameters in question and the computational time of the models belonging to a single parameter set. Theoretically, one has to find a minimum point in an 8D parameter space, where calculating a single point takes several hours. Because of this, we first reduce the number of parameters to a more manageable value.
Firstly as described by \citet{bpf-beat2002}, the  mixing length scale $\bar{\alpha}_\Lambda$ parameter is a scaling parameter in the equations; hence we can fix its value at $\alpha_\Lambda$=1.5.
Secondly, our calibration now is restricted to the RRab stars with relatively thin atmospheres compared to Cepheids. Thus the radiative cooling can be neglected by setting the parameter $\bar{\alpha}_r=0$. To check this assumption, we ran tests for this case and could not find significant differences using different $\bar{\alpha}_r$-s. This way, we have only six parameters left.

The different convective parameters change the final velocity and light curves in different ways, as we can see in Figs.~\ref{fig:bpf_parameter_change} and \ref{fig:rsp_parameter_change}. 
The two main parameters which have the most significant effect on the resulting radial velocity curves are the $\bar{\alpha}_\nu$ governing the pressure term for the eddies, and $\bar{\alpha}_d$, which describes the coupling between the turbulent and internal energy. The parameter $\bar{\alpha}_s$ that governs the source of turbulence alongside the convective flux parameter $\bar{\alpha}_c$ has little effect on the radial velocity and light curves. The turbulent pressure term $\bar{\alpha}_p$ and turbulent flux term $\bar{\alpha}_t$ has almost no effect on the observables in the case of the \bpf\  and $\bar{\alpha}_t$ has little effect in the case of the \rsp . Generally, it can be said that the \rsp\ light curves are more sensitive to the parameters than the \bpf\  ones. However, this does not mean that these three terms are negligible. Still, they are less efficiently derivable from the observable quantities. At the same time, they play a significant role in the work integral \citep{Yecko1998} and thus in the mode selection, which we are not studying here. 

The calibration is performed by calculating several parameter grids for the most prominent $\alpha$ parameters, notably $\alpha_d$ and $\alpha_\nu$. The value of $\alpha_t$, $\alpha_c$, $\alpha_s$ are loosely connected, so their fitting is done separately.

\section{Results}
\label{sec:results}

\begin{figure}
    \centering
    \includegraphics[width=\columnwidth]{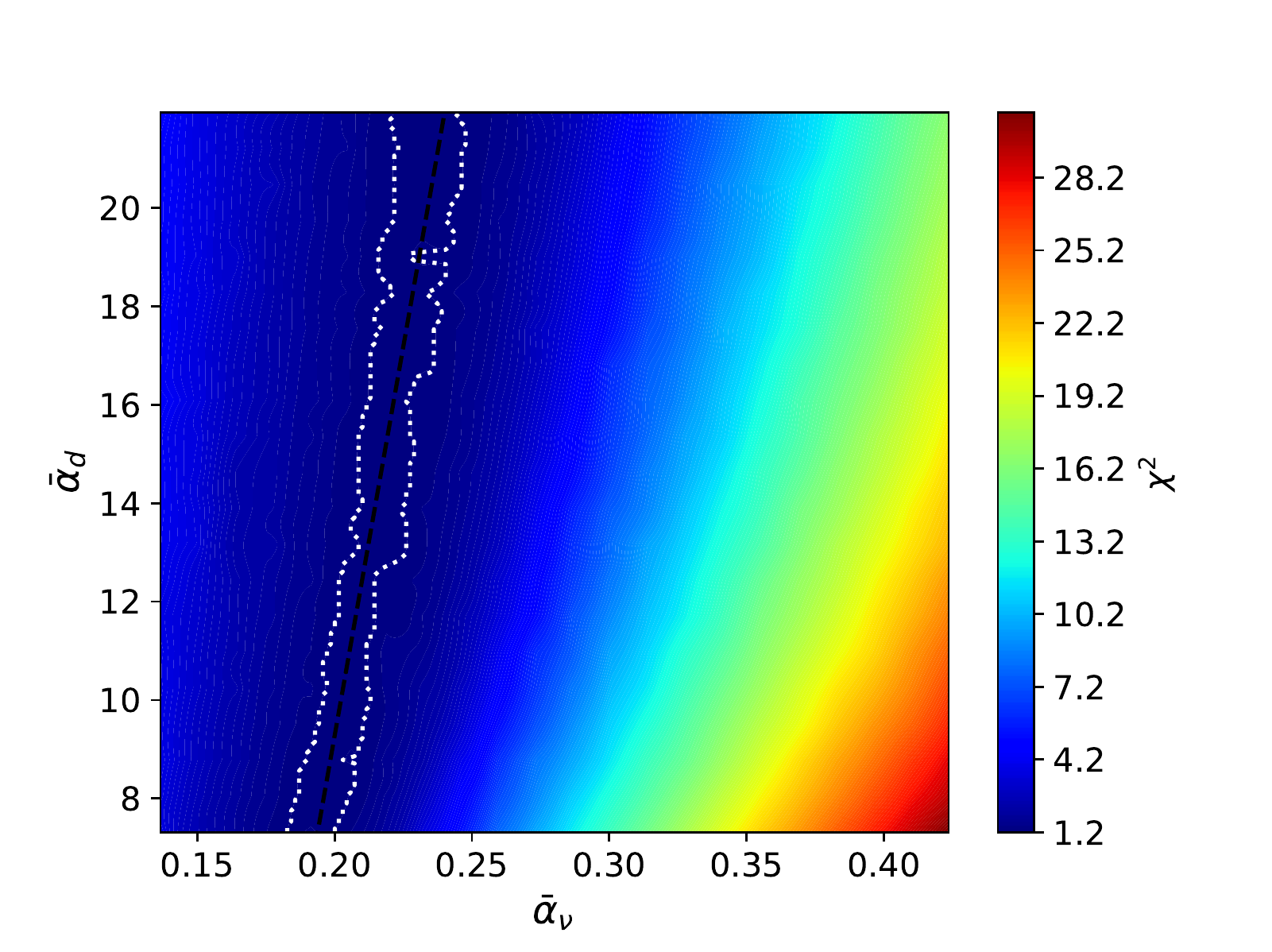}
    \caption{The structure of the parameter space in the $\bar{\alpha}_\nu$--$\bar{\alpha}_d$ plane in the case of the v036 star and the \bpf \ model. The x-axis denotes the $\bar{\alpha}_\nu$ parameter, the y-axis is the $\bar{\alpha}_d$, and the color map shows the $\chi^2$ value of the particular model. All other $\bar{\alpha}$ parameters were kept at the original \bpf\  RR Lyrae standard values \citep{bpf-beat2002}. The black dashed line is the regression line of the parameters; the white dotted curves denote the 3$\sigma$ width of the valley.}
    \label{fig:valley}
\end{figure}

The parameter space of the two prominent parameters, namely  the dissipation efficiency, $\bar\alpha_d$ and eddy viscosity scale, $\bar\alpha_\nu$, regarding the reduced $\chi^2$ values, has a valley-like structure in both models for all studied stars. We show one example of this structure in Fig.~\ref{fig:valley}.  This means that the two parameters are not independent, and one can find a linear regression between them. The valleys have a width of around $0.005$-$0.01$, so one can not make a difference between the parameters. This width will be used as one of the error sources for the fitting (see Sec. \ref{sec:errors}.).

\begin{figure}
    \centering
    \includegraphics[width=\columnwidth]{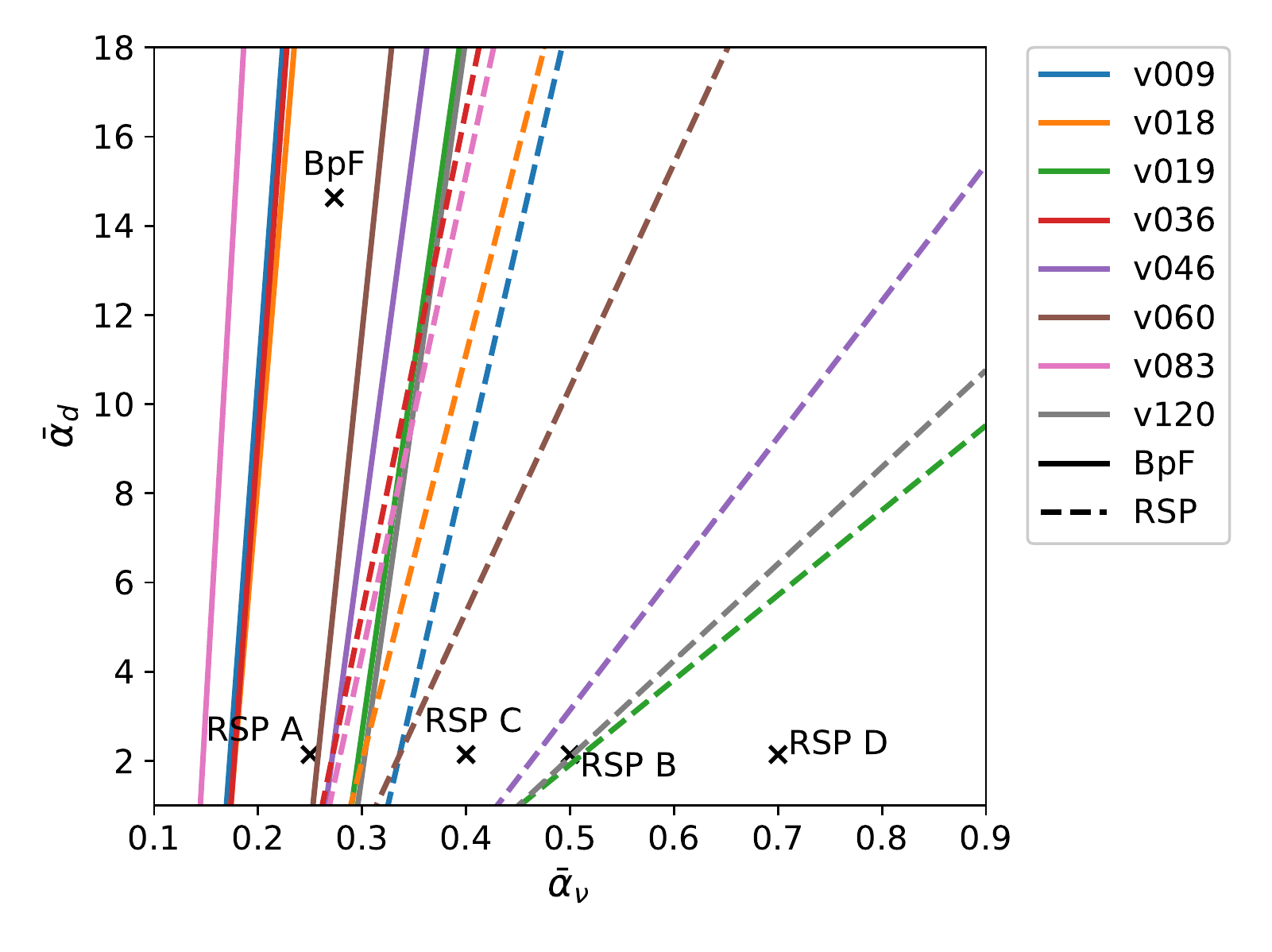}
    \caption{Regression lines on the $\bar{\alpha}_\nu$-$\bar{\alpha}_d$ plane. The colors denote the star given in the legend. Solid lines are those calculated for the \bpf \ model, and dashed lines are the \rsp \ model fits. Crosses mark the different prescriptions for the models. One can see that instead of selecting a single parameter set or single regression, the lines are spread throughout the parameter plane. The \rsp\ lines have lower steepness and shifted towards higher $\alpha_\nu$ values.}
    \label{fig:compared_regressions}
\end{figure}

We show all the regression lines in Fig.~\ref{fig:compared_regressions}. from which we can argue that there is no single independent parameter set regarding the two values. Black crosses show previous prescriptions. 

\subsection{Temperature dependence and correlations between the $\bar{\alpha}$ parameters}

The regression lines are different for the individual stars, which suggests that there should be an underlying connection between the stellar physical parameters and the parameters of the regression lines. As our sample of stars has similar [Fe/H] values,  luminosities, masses, and periods, the main difference between the stars is their temperature which covers the $T_{\rm eff} \in [6300,6800]$ regime. 
Using the stellar temperatures, we could derive a regression between the $\bar{\alpha}_d = a+b\bar{\alpha}_\nu$  line parameters $a$,$b$ and $T_{\rm eff}$. For convenience, we switch to use $\bar{\alpha}_\nu(\bar{\alpha_d})$ functions because  we have better prescriptions for $\bar{\alpha}_d$,
thus we use $-a/b$ for the intercept and $b^{-1}$ for the slope and the logarithm of the effective temperature.
The result can be seen in Fig.~\ref{fig:slope_fit} and Fig.~\ref{fig:intercept_fit}.
The following subsections describe the results separately for the two numerical codes.

\subsubsection{Budapest-Florida code results}
\label{sec:bpf-results}

\begin{figure*}
    \centering
    \includegraphics[width=\textwidth]{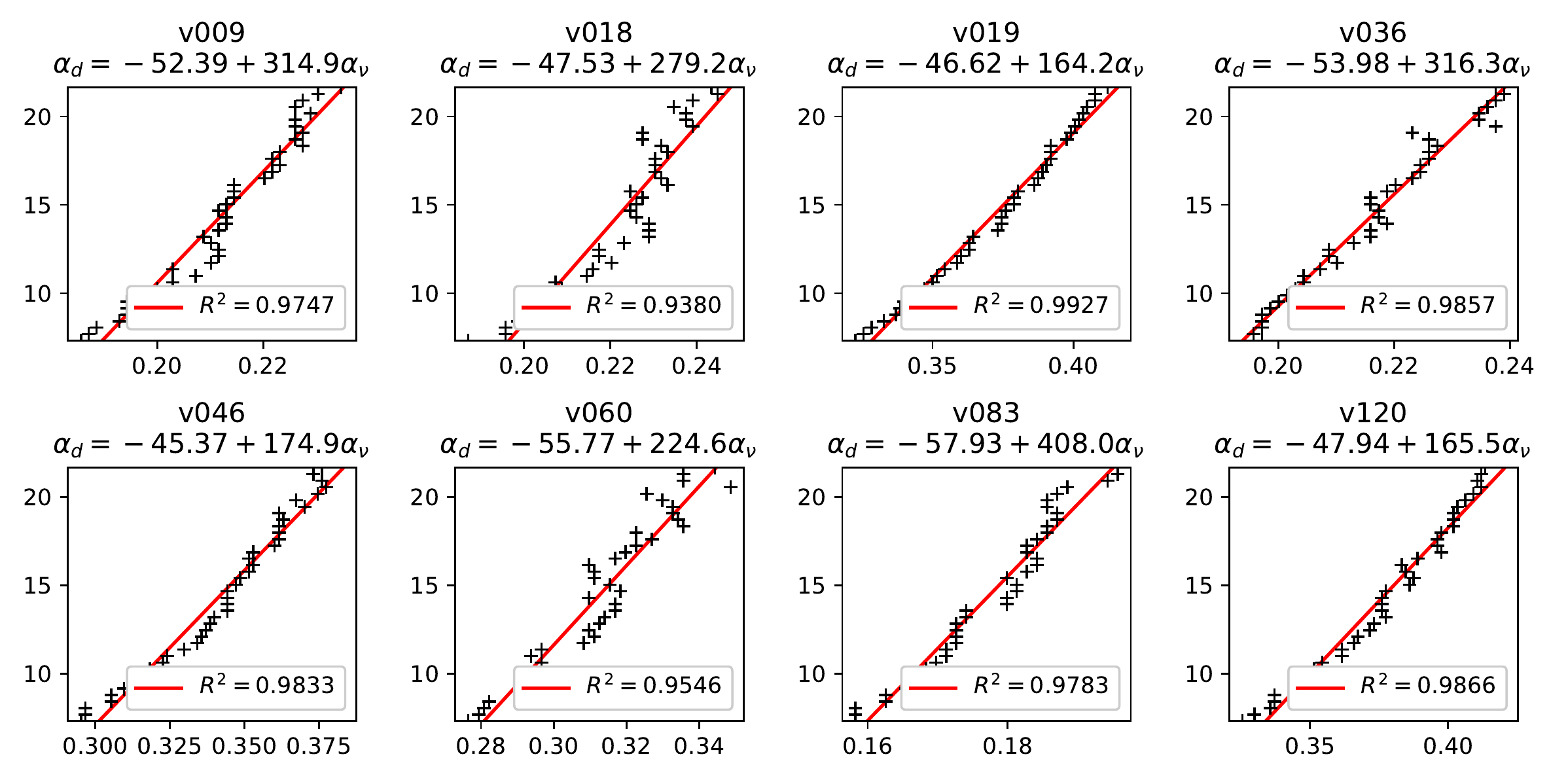}
    \caption{Linear regressions between $\alpha_\nu$ and $\alpha_d$ for each star in the case of the \bpf\ model. Best fitting $\alpha_d$ for given $\alpha_\nu$ values (black crosses). The red line is the linear regression for each star.$R^2$ values shown in the legends, and the parameters can also be read in Table. \ref{tab:regression_coefficients}.}
    \label{fig:regression_bpf}
\end{figure*}

The individual regression lines for each  star can be seen in Fig.~\ref{fig:regression_bpf}, and the derived parameters can be found in Table~\ref{tab:regression_coefficients}; the regression is weaker when $\bar{\alpha}_d \lesssim 9$.
Also, in a statistical sense, there is no difference between the different $\bar{\alpha}_\nu-\bar{\alpha}_d$ pairs, while the best fitting values are around  $9.6-11.6$. The derived connection between $\bar{\alpha}_\nu-\bar{\alpha}_d-T_{\rm eff}$ is:

\begin{multline}
\label{eq:a_d_a_nu_t_bpf}
    \bar{\alpha}_\nu = (-5.4\pm 0.36) \log T_{\textrm{eff}} - (0.12 \pm 0.01) \log T_{\textrm{eff}} \bar{\alpha}_d\\
    + (0.495 \pm 0.05) \bar{\alpha}_d + (20.83 \pm 1.3)
\end{multline}

where the rms of the fit is $0.0112$. 

The  turbulent flux efficiency, $\bar{\alpha}_t$ parameter has minimal effect in this code but improves the RV fits at around $\bar{\alpha}_t\approx0.25$, so we use the original prescription $\bar{\alpha}_t=0.2733$. 

The  turbulent source parameter, $\bar{\alpha}_s$ and  convective flux efficiency, $\bar{\alpha}_c$ were fitted separately, and we found two groups for the parameters where we have a good fit. The results for each star can be found in Table \ref{tab:parameters}. The first group, let's call it set $\mathfrak{A}$, consists of an almost zero $\bar{\alpha}_s$ and an $\bar{\alpha}_c \approx 0.3$; in the second group (set  $\mathfrak{B}$)  the two parameters are of the same order (0.17 and 0.22). 
In the case of set $\mathfrak{B}$ the  turbulent source parameter ( $\bar{\alpha}_s$ ) has a larger standard deviation. It shows a weak temperature dependence, but statistically, it is not significant ($R^2 < 0.5$). . More model computations would be necessary to establish such a relation if it exists.

For the final parameter sets, we took the averages and standard deviations from Table \ref{tab:parameters}, and these results are in Table \ref{tab:a_c_a_s_sets}.

\begin{figure*}
    \centering
    \includegraphics[width=\textwidth]{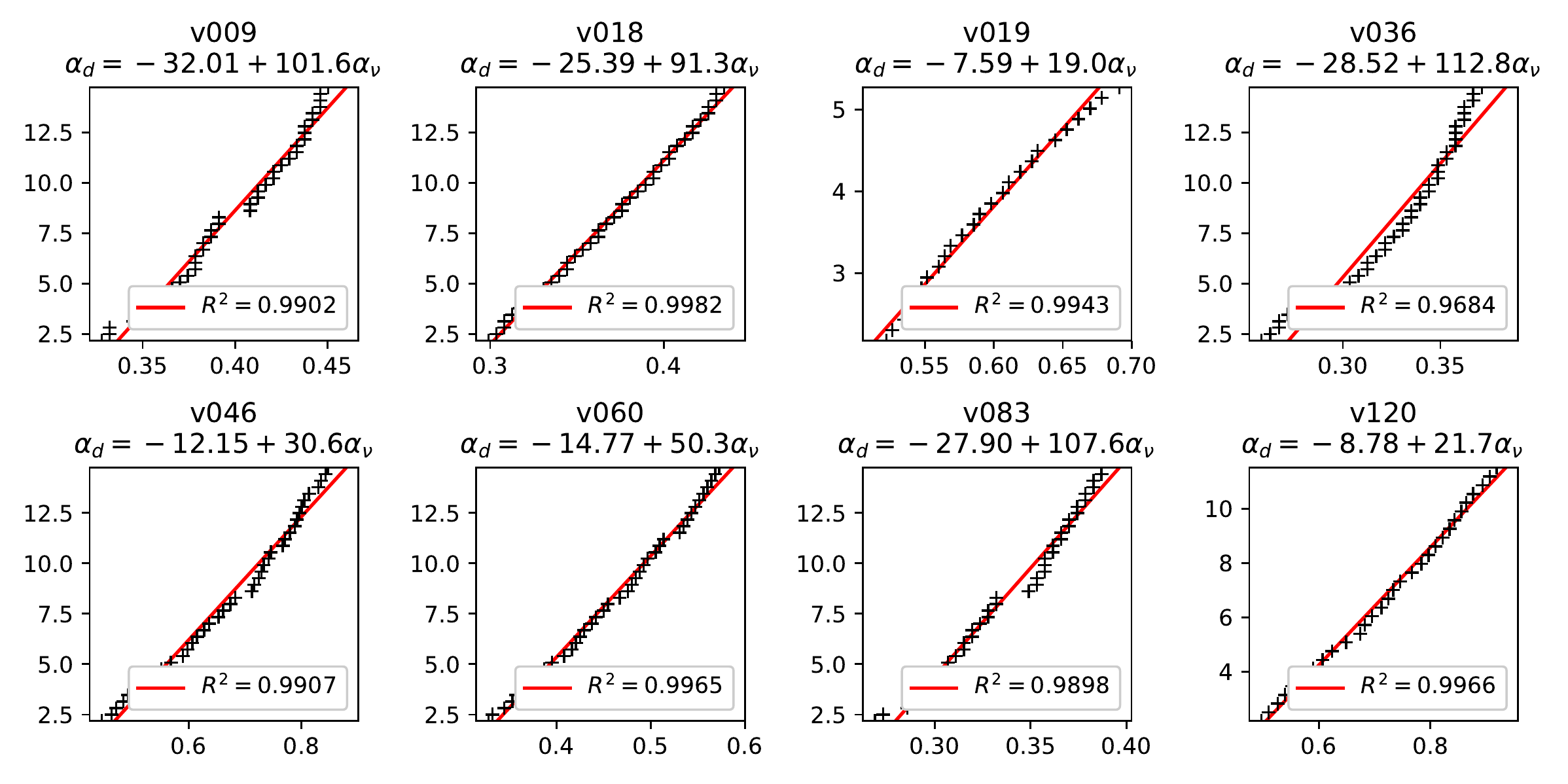}
    \caption{The same as Fig. \ref{fig:regression_bpf}. but for the \rsp\ models.}
    \label{fig:regression_rsp}
\end{figure*}

\begin{table}
    \centering
    \caption{Parameters of linear regression between $\bar{\alpha}_\nu$ and $\bar{\alpha}_d$. The formula used is: $\bar{\alpha}_d = a + b \bar{\alpha}_\nu$.}
    \begin{tabular}{c|c|c|c|c|c|c}
            & \multicolumn{3}{|c|}{BpF} & \multicolumn{3}{|c|}{MESA RSP} \\
        Star & $a$ & $b$ & $R^2$ & $a$ & $b$ & $R^2$\\
        \hline
        v009 & -37.86 & 241.30 & 0.9786&  -32.01 & 101.61 & 0.9804\\
        v036 & -53.98 & 316.27 & 0.9857 & -28.52 & 112.80 & 0.9380\\
        v018 & -34.61 & 215.77 & 0.8661 & -25.39 & 91.28 & 0.9964\\
        v019 & -46.62 & 164.23 & 0.9855 & -7.59  & 19.01 & 0.9943\\
        v046 & -45.37 & 174.90 & 0.9833 & -12.15 & 30.58 & 0.9814\\
        v060 & -55.77 & 224.64 & 0.9546 & -14.77 & 50.28 & 0.9965\\
        v083 & -57.93 & 407.96 & 0.9571 & -27.90 & 107.61 & 0.9898\\
        v120 & -47.94 & 165.45 & 0.9733 & -8.78 & 21.71 & 0.9966\\
    \end{tabular}

    \label{tab:regression_coefficients}
\end{table}

\subsubsection{RSP code results}

\begin{figure}
    \centering
    \includegraphics[width=\columnwidth]{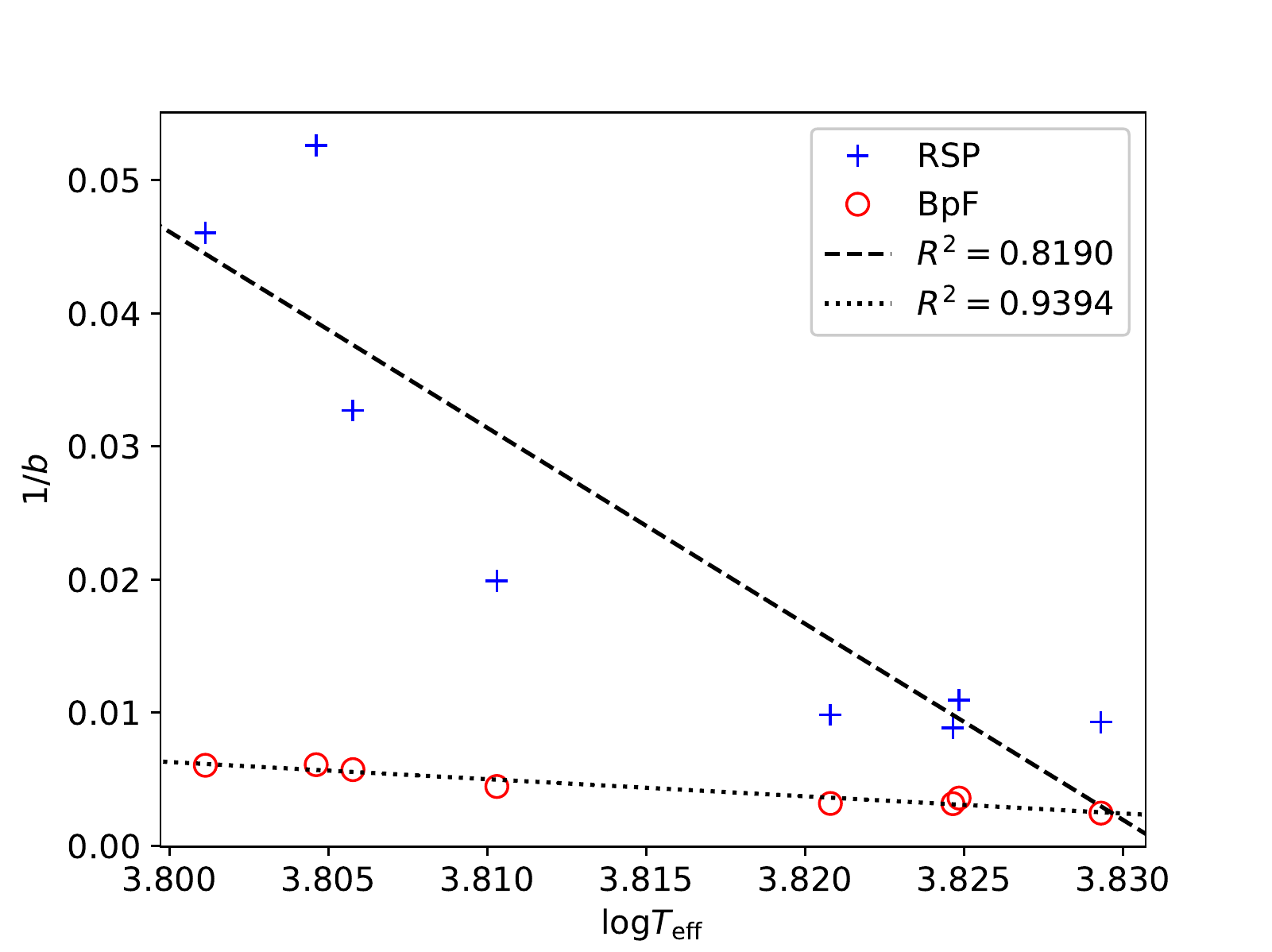}
    \caption{Slope parameter of the $\bar{\alpha}_\nu(\bar{\alpha}_d)$ function vs the logarithm of the effective temperatures. Blue crosses are the values from the \rsp\ fit, and red circles are the values of the \bpf\ fit. Black dashed and dotted lines are the linear regressions for the \rsp\ and \bpf\ slopes, respectively.}
    \label{fig:slope_fit}
\end{figure}

\begin{figure}
    \centering
    \includegraphics[width=\columnwidth]{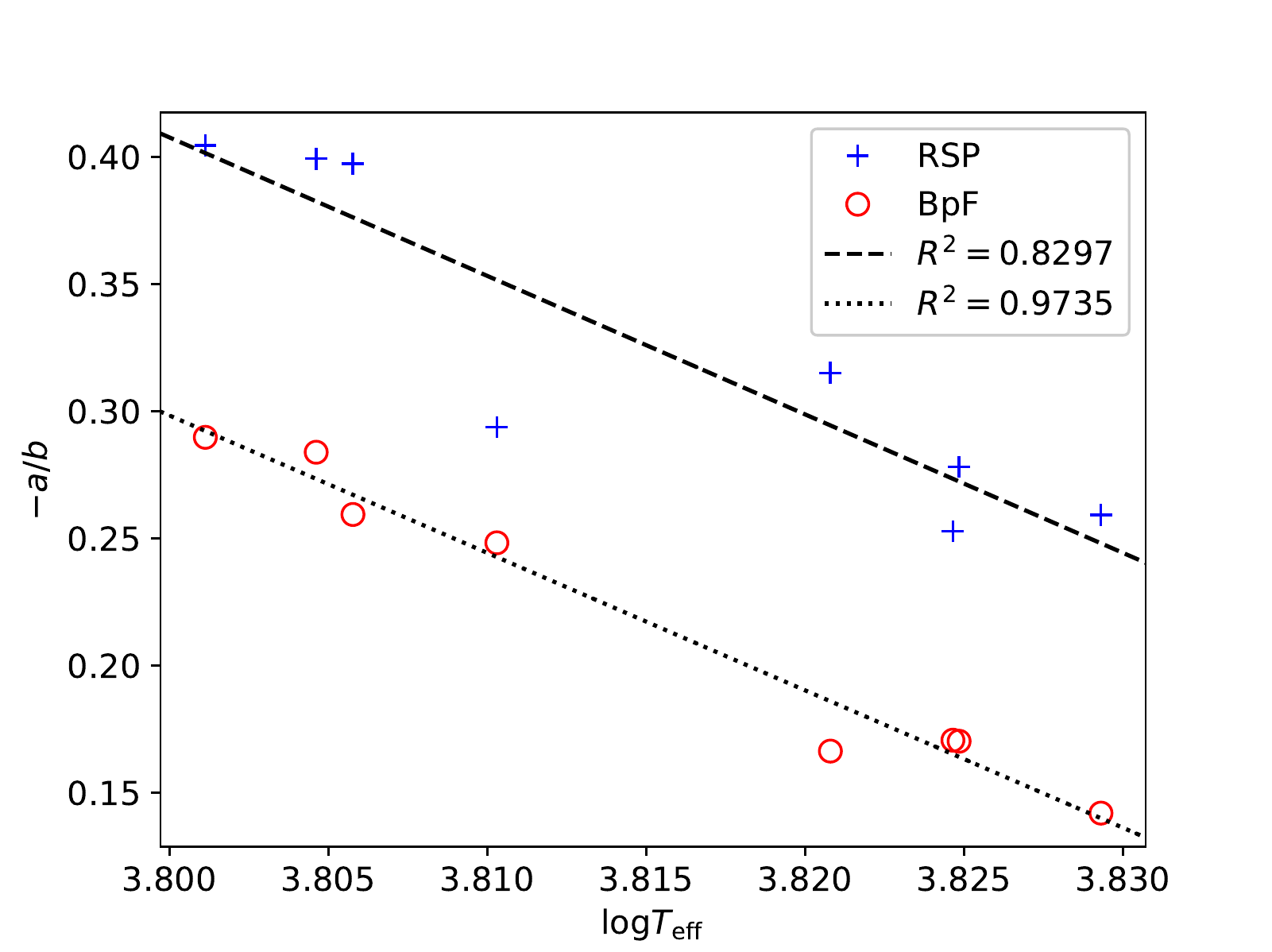}
    \caption{The same as Fig. \ref{fig:slope_fit} but for the interception parameters of the $\bar{\alpha}_\nu(\bar{\alpha}_d)$ function.}
    \label{fig:intercept_fit}
\end{figure}

In the case of the \rsp\ we show the  eddy viscosity vs. dissipation efficiency, i.e. $\bar{\alpha}_\nu$--$\bar{\alpha}_d$ regression lines in Fig. \ref{fig:regression_rsp}, and the derived parameters are in Table \ref{tab:regression_coefficients}. Here the regression is maintained at low $\bar{\alpha}_d$, as well. The fitted lines  have smaller slopes and the parameters depend on the effective temperature. These results are shown in Fig.~\ref{fig:slope_fit}. and \ref{fig:intercept_fit}. 

The final $\bar{\alpha}_\nu$-$T_\textrm{eff}$-$\bar{\alpha}_d$ relation for this case is the following:

\begin{multline}
\label{eq_a_d_a_nu_t_rsp}
    \bar{\alpha}_\nu = (-5.4\pm 1) \log T_{\textrm{eff}} - (1.47 \pm 0.28) \log T_{\textrm{eff}} \alpha_d \\
    + (5.6 \pm 1.1) \alpha_d + (21.1 \pm 3.8)
\end{multline}
where the rms of the fit is $0.0377$.

\begin{table}
	\centering
	\caption{Table of fitting $\alpha_s$ and $\alpha_c$  parameters for all M3 calibrator stars. For each star, we show the first two best-fit parameter pairs. In the case of the \bpf\ one can find two distinct parameter sets, while the differences in the case of the \rsp \ are insignificant.}
	\label{tab:parameters}
	\begin{tabular}{lcccccr}
		\hline
		\multirow{2}{*}{star} & \multicolumn{3}{c}{BpF} & \multicolumn{3}{c}{RSP} \\
		 & $\bar{\alpha}_c$ & $\bar{\alpha}_s$ & $\chi^2$ & $\bar{\alpha}_c$ & $\bar{\alpha}_s$ & $\chi^2$\\
		\hline
		\multirow{2}{*}{v009}  & 0.3096 & 0.0060 & 1.0240 & 0.2399 & 0.1611 & 1.4588 \\
                               & 0.1506 & 0.1721 & 1.0913 & 0.3299 & 0.1611 & 1.6037 \\
        \multirow{2}{*}{v018}  & 0.3431 & 0.0007 & 0.8613 & 0.2399 & 0.2964 & 0.8236 \\
                               & 0.2176 & 0.4201 & 1.1799 & 0.2250 & 0.4017 & 0.8555 \\
        \multirow{2}{*}{v019}  & 0.3263 & 0.0027 & 0.5538 & 0.3149 & 0.3566 & 0.6223 \\
                               & 0.1506 & 0.2151 & 0.6758 & 0.2549 & 0.3716 & 0.6399 \\
        \multirow{2}{*}{v036}  & 0.2929 & 0.0027 & 1.1070 & 0.2399 & 0.2964 & 1.3013 \\
                               & 0.1506 & 0.1513 & 1.1484 & 0.3224 & 0.2814 & 1.4225 \\
        \multirow{2}{*}{v046}  & 0.3263 & 0.0029 & 0.8425 & 0.2474 & 0.4017 & 0.6200 \\
                               & 0.1590 & 0.2332 & 1.0042 & 0.3149 & 0.4017 & 0.6398 \\
        \multirow{2}{*}{v060}  & 0.2761 & 0.0060 & 0.9867 & 0.2549 & 0.4017 & 0.5165 \\
                               & 0.1924 & 0.1721 & 1.0018 & 0.2474 & 0.3115 & 0.5203 \\
        \multirow{2}{*}{v083}  & 0.3012 & 0.0060 & 1.2160 & 0.2399 & 0.2814 & 1.4906 \\
                               & 0.1506 & 0.1721 & 1.3231 & 0.3299 & 0.2664 & 1.6566 \\
        \multirow{2}{*}{v120}  & 0.3263 & 0.0027 & 0.5846 & 0.3074 & 0.5371 & 0.7319 \\
                               & 0.1422 & 0.2414 & 0.7519 & 0.2924 & 0.7326 & 0.8193 \\
		\hline
	\end{tabular}
\end{table}

\begin{table}
    \centering
     \caption{The final parameter sets of $\bar{\alpha}_s$ and $\bar{\alpha}_c$, taken as averages from Table \ref{tab:parameters}.}
    \begin{tabular}{c|c|c}
        set & $\bar{\alpha}_c$ & $\bar{\alpha}_s$ \\
        \hline 
        \bpf \ $\mathfrak{A}$ & $0.31 \pm 0.02$ & $0.004 \pm 0.002$ \\
        \bpf \ $\mathfrak{B}$ & $0.17 \pm 0.02$ & $0.22 \pm 0.08$ \\
        \rsp     & $0.27 \pm 0.03$ & $0.31 \pm 0.07$ 
    \end{tabular}
   
    \label{tab:a_c_a_s_sets}
\end{table}

The  turbulent flux efficiency parameter, $\bar{\alpha}_t$ has the best-fit value around $0.24\pm0.03$ for all stars. Using a larger value does not have any additional effect on the RV or light curves. 

In this case, the fitting for $\bar{\alpha}_c$  (convective flux efficiency)  and $\bar{\alpha}_s$  (efficiency of turbulent source) shows a single set. Interestingly, there is a parameter set where $\alpha_c$ is close to zero, but it never fits better than the set of values. One can also see from Fig.~\ref{fig:rsp_parameter_change} that an $\bar{\alpha}_c$ parameter that low would increase the amplitude in the luminosity too much. The two best-fitting value pairs of this fit can be seen in Table \ref{tab:parameters}.

\subsection{Sources of errors}
\label{sec:errors}

\begin{figure}
    \centering
    \includegraphics[width=\columnwidth]{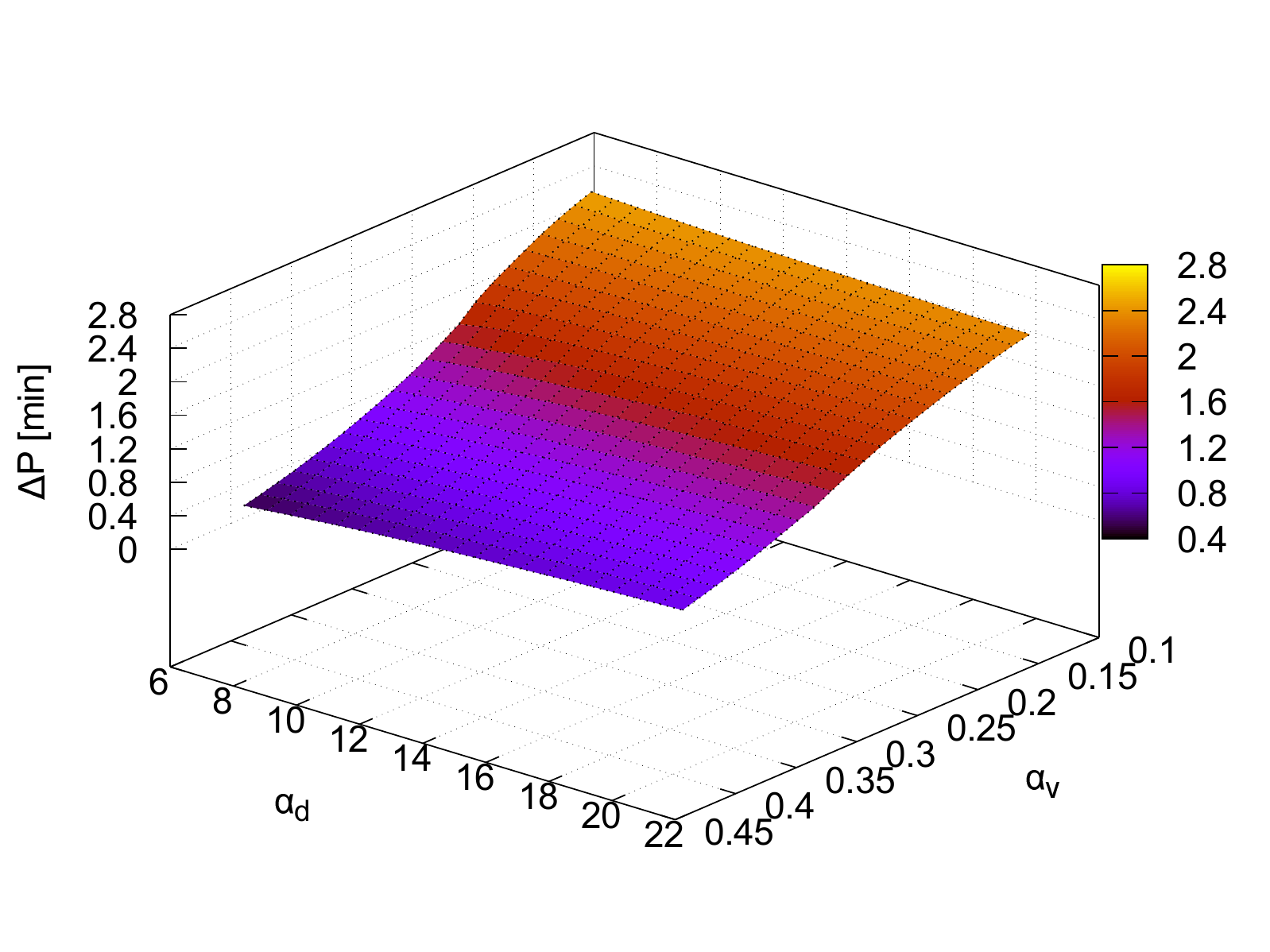}
    \caption{Deviation in the pulsation period between the linear and nonlinear calculations vs. the convective parameters $\bar{\alpha}_d$ and $\bar{\alpha}_n$.}
    \label{fig:non-linear_effect}
\end{figure}

There are multiple sources of errors in the parameter estimations. The first is that the nonlinear models' periods deviate from the linear models by a few minutes, which affects the mass estimation; the period difference between linear and nonlinear models is around $0.001$ day  which value is weakly dependent on the chosen $\bar{\alpha}$ parameters. The deviation effect of the $\bar{\alpha}_\nu$ and $\bar{\alpha}_d$ parameters is the strongest, we show this in Fig.~\ref{fig:non-linear_effect}. This phenomenon introduces
an error of about $0.002\ M_\odot$ in static model mass. This has a negligible ($<0.001$) effect on the final  $\bar{\alpha}$ parameters.
 The second source is the uncertainty of the p-factor. This is included in the error of individual points of the observed RV curve by using Gaussian error propagation.
 Another source is that the observational uncertainties give 3--5 $L_\odot$ error in calculating the luminosities and, therefore,  $0.05$-$0.1$ $M_\odot$ in the mass estimation. This translates in the final fit into an estimated error value of $0.005$ in the $\bar{\alpha}$s , due to the mass of the model is determined through the period.
The valley-like structures have a width of around $0.005$--$0.01$ in the case of the \bpf\ and $0.01$--$0.04$ in the case of \rsp.    These values are very similar to the rms of the fitted formulae given by Eq. (\ref{eq:a_d_a_nu_t_bpf}) and (\ref{eq_a_d_a_nu_t_rsp}) 
hence we handle them together. This way the overall statistical uncertainties of the $\bar{\alpha}_\nu$ parameter is $0.015$ in the case of the \bpf\ and $0.04$ in the case of the \rsp, respectively.

The systematic errors introduced into the calculations may come from assuming incorrect luminosities or values of $X$ and $Z$. Test calculations were done to test the magnitude of the effect coming from the uncertainties of these parameters for each hydrocode. Using different luminosities, the regression parameters $a$ and $b$ are affected by 6\% and 2\%, respectively, decreasing these values. This means around $0.01$ decrease in the final fit, which is in the range of the fit's initial error. This error can also be included in the uncertainties of the $L$.  The systematic errors that the metallicity and helium content can introduce are $0.01$ in the case of the \bpf\  and $0.03$ in the case of the \rsp .

These errors increase the systematic uncertainty of the final parameter sets of $\bar{\alpha}_\nu$ to $0.035$ for \rsp\  and $0.015$ for the \bpf . This value is in the same range as the original valley width, meaning that using our estimation for a given $\bar{\alpha}_d$ and $T_{\rm eff}$ for $\bar{\alpha}_\nu$ gives a reasonably good fit for the RV curve.

\subsection{Comparison of the \bpf\ and \rsp\ codes} 
\label{sec:results-compare}

We found several differences between the models that are not expected by solely looking at the equations. This can be due to the nonlinear nature of the problem, which can enhance the consequences of the differences in the underlying physical models. The $\bar{\alpha}_d(\bar{\alpha}_\nu)$ functions have lower slopes in the case of the \rsp\ models showing a more subtle effect of the $\alpha_\nu$ parameter on the RV curves. The finalized versions of the $\bar{\alpha}_\nu(\bar{\alpha}_d,T_{\rm eff})$ functions have a similar dependence on the temperature, but  the \rsp\ models have stronger $\bar{\alpha}_d$ dependence. In terms of other parameters, the \bpf\ $\mathfrak{B}$ set for $\bar{\alpha}_c$ and $\bar{\alpha}_s$ (convective flux  and turbulent source efficiency) are about $2/3$ of the \rsp\ values . This suggests that the convective effects are up-scaled somewhat in the \bpf , thus, we need lower parameters for getting a fit to the RVs. 

\subsubsection{Final fitted RV curves vs. light curves}
\label{sec:comp-fits}

\begin{figure*}
    \centering
    \includegraphics[width=\textwidth]{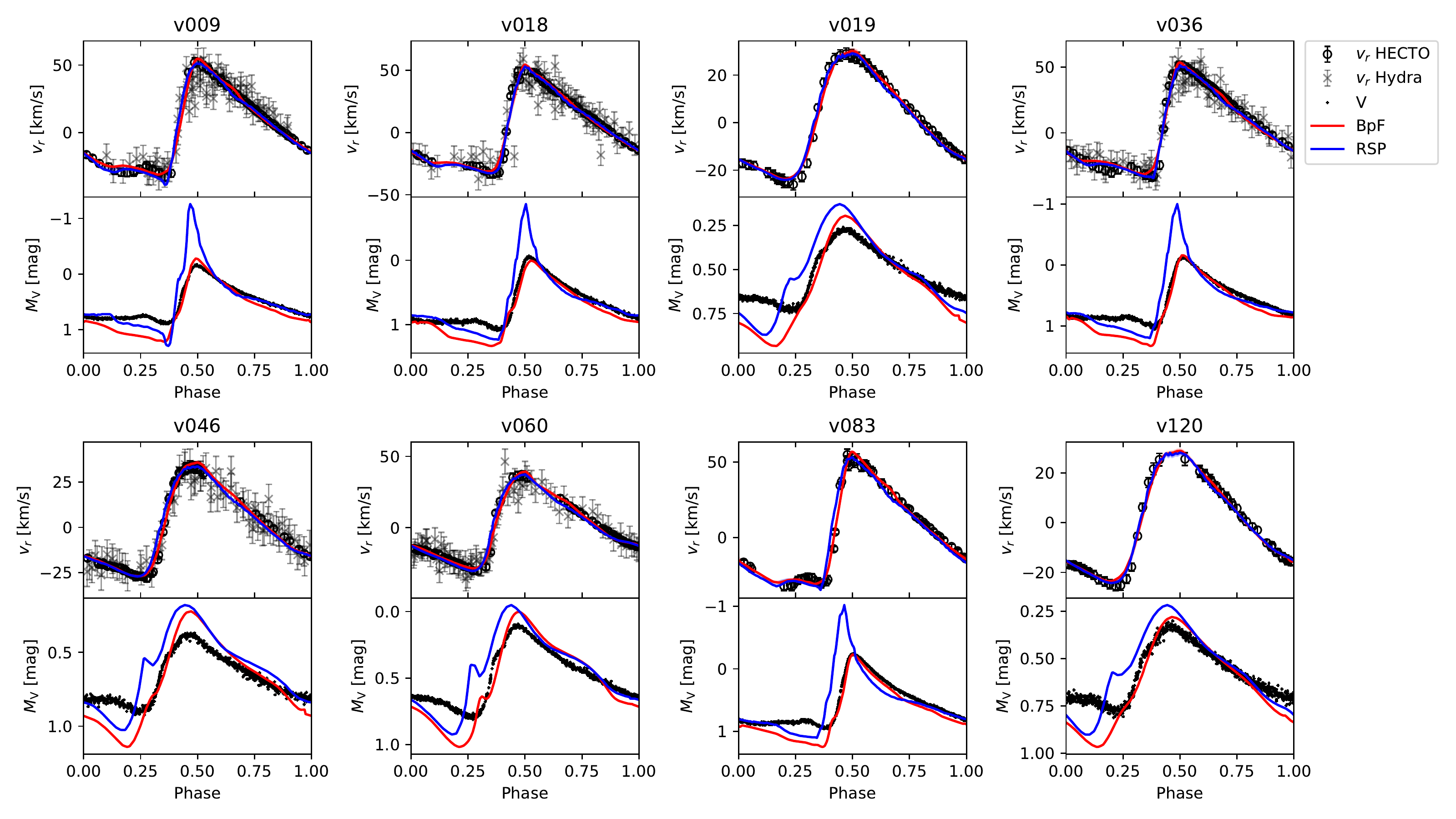}
    \caption{Models fitted to the radial velocities for each star. Upper panels show the radial velocity curve: crosses are measurements of the \hydra\, and empty circles are measurements of \hecto . The red line is the \bpf\ model, and the blue line is the \rsp\ model. The lower panels show the same models' light curves in V magnitude, black dots are the observations, and the lines have the same meaning as in the upper panel. The light curves were calculated by using the $BC_V$ correction of \citet{DorfiFeucthinger1999} on the bolometric light curves at $\tau=2/3$ optical depth. One can see that while RV fits the amplitude  in most cases, only missing the bumps and humps around the minima sometimes, the light curves strongly differ for the two models. The \bpf\ models can fit the maxima and the ascending branch but have very deep minima. Meanwhile, the \rsp\ shows an overall larger amplitude, and the bumps appear at the wrong phases.}
    \label{fig:best_fit_curves}
\end{figure*}

 In this subsection, we compare the specified models' radial velocity and absolute $M_{{\rm V}}$ magnitude light curves. The models are based on their best-fit parameters:  individual $\bar{\alpha}_c$ and $\bar{\alpha}_s$ values are taken from Table~\ref{tab:parameters}. choosing set $\mathfrak{B}$ in the case of \bpf. $\bar{\alpha}_d$ is set to 10.6 for the \bpf\ and $2.17$ for the \rsp\ codes, respectively. (We discuss these particular $\bar{\alpha}_d$ choices in Section \ref{sec:recommendend_alpha}), $\bar{\alpha}_{\nu}$ is calculated based on the individual regression lines (see Table~\ref{tab:regression_coefficients}). $\bar{\alpha}_p$, turbulent pressure scale is kept at $2/3$, and  the turbulent flux efficiency, $\alpha_t=0.24$ for \rsp, and $0.2733$ for the \bpf\ \citep[the presribed value given by][]{bpf-beat2002}. As the models are calibrated based on the observed RV curves, and in the absence of any atmosphere models,  the light curves can be only approximated (that is done by  using the bolometric correction formula of \citet{DorfiFeucthinger1999} for the $M_{\rm V}$), it is reasonable not to expect a perfect fit of the light curve, rather agreement with observations in the light curve morphology or the amplitude. Hence we illustrate only the light curves of \rsp\ and \bpf\ models.

In Fig. \ref{fig:best_fit_curves}. we can see the fitted radial velocity curves (upper panels) and the V light curves (lower panels) of the corresponding models for each star. We can see that the best-fitting models fit well the RV curves, but the corresponding light curves deviate substantially. 

Albeit both models fit poorly the $M_{\rm V}$ light curves, the \bpf\ seems to perform better. The light curve maxima are fitted well, but the light variation minima predicted by this numerical code are too low, even lower than in the \rsp\ case. The \rsp\, on the other hand, has larger amplitudes accompanied by strong bumps, which are not as prominent in the observed light curves. There is a trend in the amplitudes, too: the lower-temperature models fit better than the higher-temperature ones, with an amplitude of about twice the observed ones.

To investigate the origins of these differences and effects, we made an in-depth analysis by comparing the internal structure variations in stellar models provided by the two pulsation codes. We describe the results of this analysis in Sections~\ref{sec:comp-static},~\ref{sec:comp-limitcycle}, and discuss the possible reasons and consequences in Sectiond~\ref{sec:discuss-diff}.

\subsubsection{Differences in the static models}
\label{sec:comp-static}

\begin{figure*}
    \centering
    \includegraphics[width=\textwidth]{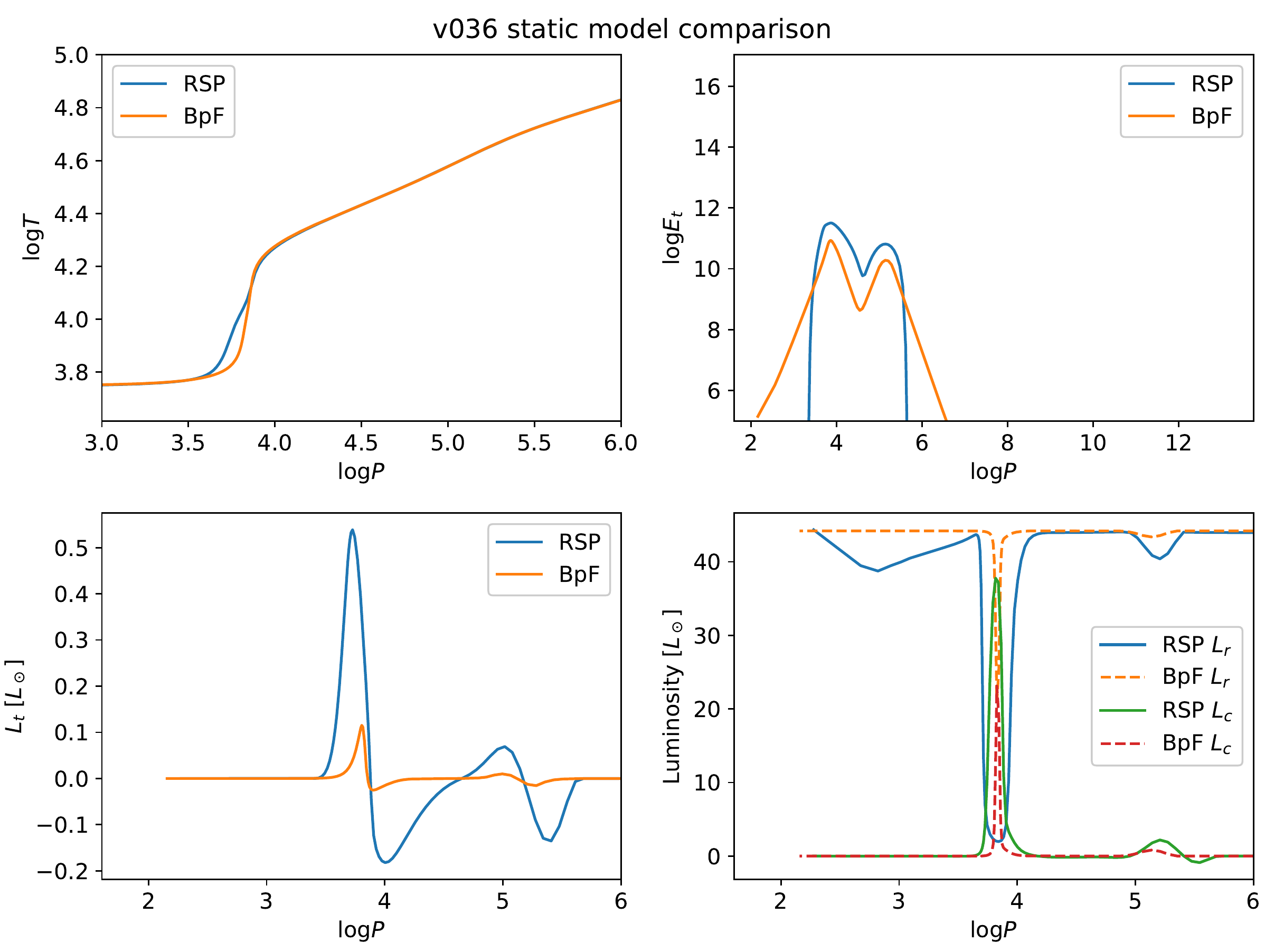}
    \caption{Comparison of the two static models for the star v036. The upper left panel shows the $\log T$ profile vs. the $\log P$. Blue and orange lines are the \rsp\ and \bpf\ models, respectively. The upper right panel is similar but for the specific turbulent energy $e_t$. In the lower left panel, we can see the differences in the turbulent luminosity $L_t$ profile. The lower right panel shows the radiative and convective luminosities. The blue and green solid lines are \rsp\ $L_r$ and $L_c$ curves, respectively. The dashed orange line is the \bpf\ $L_r$, and the dashed red line is the \bpf\ $L_c$. For details, see the text.}
    \label{fig:v036_static}
\end{figure*}

As we mentioned, these model calculations work in a way that they build up a static envelope model giving it an initial velocity profile; hence this first static model is essential. We directly compare the two models for the star v036 in Fig. \ref{fig:v036_static}. The Figure shows the temperature profiles (upper left panel), the specific turbulent energy profile $e_t$ (upper right panel), the turbulent luminosity, $L_t$ (lower left panel), and the convective and radiative luminosity profiles $L_r$,$L_c$ (lower right panel). All profiles are plotted as a function of the thermodynamic pressure.

The first difference is that the \rsp\ has a somewhat wider H-ionization zone with a less steep temperature gradient. As convection has a widening effect and we have seen from the parameters that the convective parameters are larger in the \rsp,\ this phenomenon can be attributed to the stronger convection. In the upper right panel, we can see that in the \rsp\ model, the edges of the convective region are very sharp. In contrast, the \bpf\ has a more diffuse energy profile, which physical interpretation is a pronounced overshooting. The phenomenon was described previously by \citet{lengyel,lengyel2}, arguing that this feature causes the double mode pulsations in the \bpf\ models. 

The other main difference is that the \rsp\ has a powerful turbulent luminosity, breaking the energy equation for the static models. If we look at Eqs. (\ref{eq:energy}) and (\ref{eq:turbulent_energy}) and set the left-hand sides and $\mathcal{D}_E$ zero, we can see that the divergences of the $(F_c+F_r)$ and $F_t$ terms should be equal with opposite signs, and their absolute value should be equal to the coupling term. This is true for the \bpf\, but in the case of the \rsp\, it does not happen. The problem here is that the $L_c$ curve in the H-ionization region is much thinner than the $L_r$ curve. There is also an artificial hump near the surface in the $L_r$ profile. These problems are present when one sets $\bar{\alpha}_t=0$ meaning that $F_t=0$, so $\vdiv (F_t) =\mathcal{C} = 0 = - \vdiv (F_c+ F_r)$. This affects the energy balance and brings additional energy into the system. It seems to be rather an inaccuracy in the \rsp\ code itself than a consequence of the physical model, but this phenomenon requires further investigation. 

\subsubsection{Comparison of the limit cycles}
\label{sec:comp-limitcycle}

\begin{figure*}
    \centering
    \includegraphics[width=\textwidth]{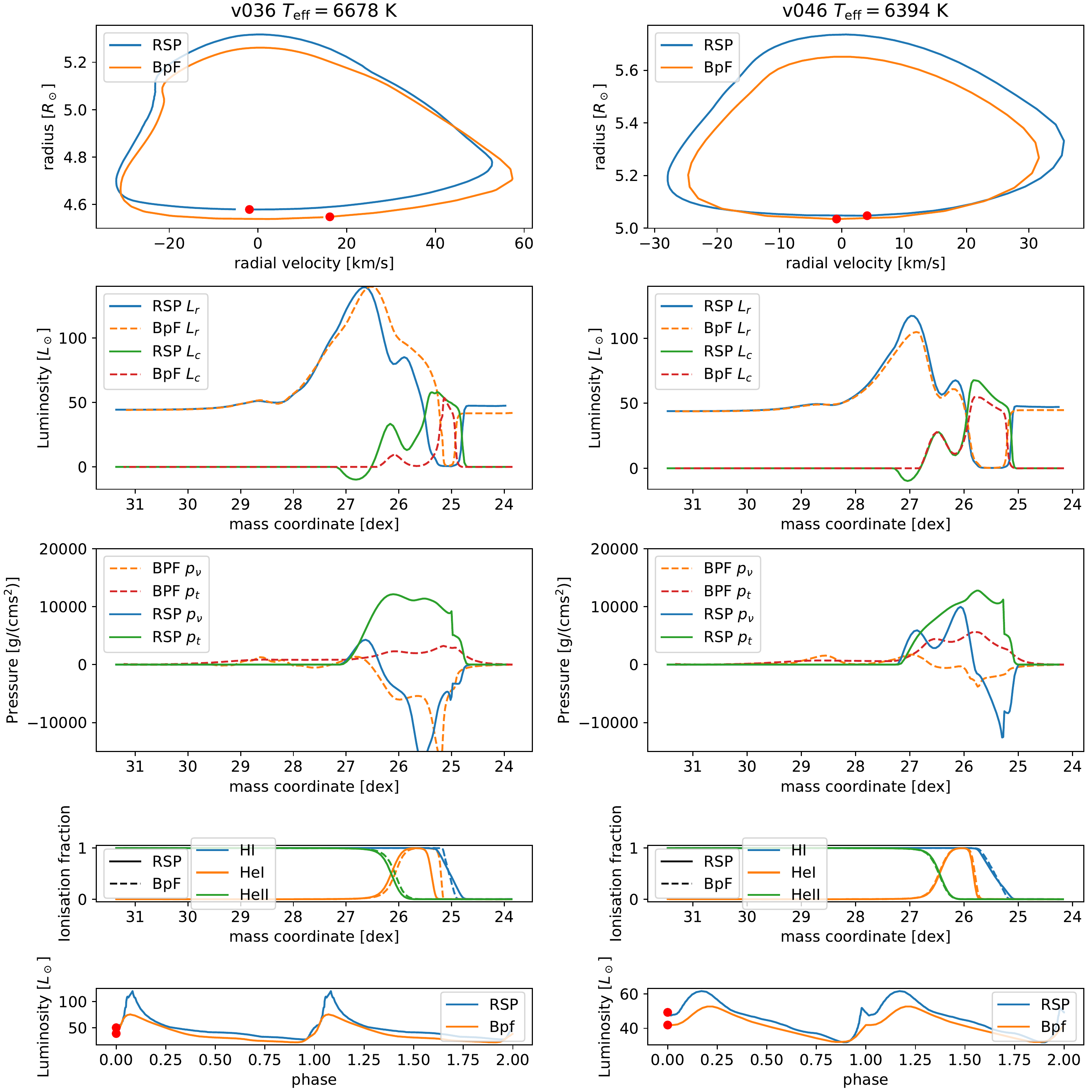}
    \caption{Snapshot of the video attached to this paper, at the minimum radius phase. Two stars from our sample are shown. On the left hand side is the M3 RRab v036 ($T_{\rm eff} = 6678$ K), while on the right hand side v046 ($T_\textrm{eff} = 6394$ K) is shown. The first panel from the top shows the radial velocity ($v_r$) -- radius ($r$) phase space and the limit cycle of the two models. Blue is the \rsp\, orange is the \bpf\  limit cycle. We show radial profiles of various quantities from the second to the fourth panels. Mass coordinates calculated from the surface are shown on the x-axis. The second panel shows the radiative and convective luminosity profiles: \bpf\ $L_c$ is the red dashed line, $L_r$ is the orange dashed line, \rsp\ $Lc$ is the solid green line, $L_r$ is the solid blue line. The third panel shows the eddy viscosity pressure, $p_\nu$ (solid blue and dashed orange is \rsp\ and \bpf\, respectively), and the turbulent pressure profile, $p_t$ (\rsp\ is the solid green,  \bpf\ is the dashed red line). The fourth panel shows the ionization fractions of H{\scshape I} (blue), He{\scshape I} (orange), and He{\scshape II} (green) vs. the mass coordinate. Solid and dashed lines are the \rsp\ and \bpf\ models, respectively. The bottom panel shows phase folded light curves calculated at $\tau=1/2$ for two periods in solar luminosities. The red circles in the top and bottom panels show the actual positions of the models in time and phase space.
    One can see the prominent differences between the models: the slightly larger radii of the \rsp\ (top panel), the presence of negative convective luminosity in the case of the \rsp\ (second panel), the presence of $p_t$ and $p_\nu$ in the deeper regions of the \bpf\ models (third panel), and the large luminosity peak excess of the \rsp\ (bottom panel).}
    \label{fig:snapshot0}
\end{figure*}%

\begin{figure*}
    \centering
    \includegraphics[width=\textwidth]{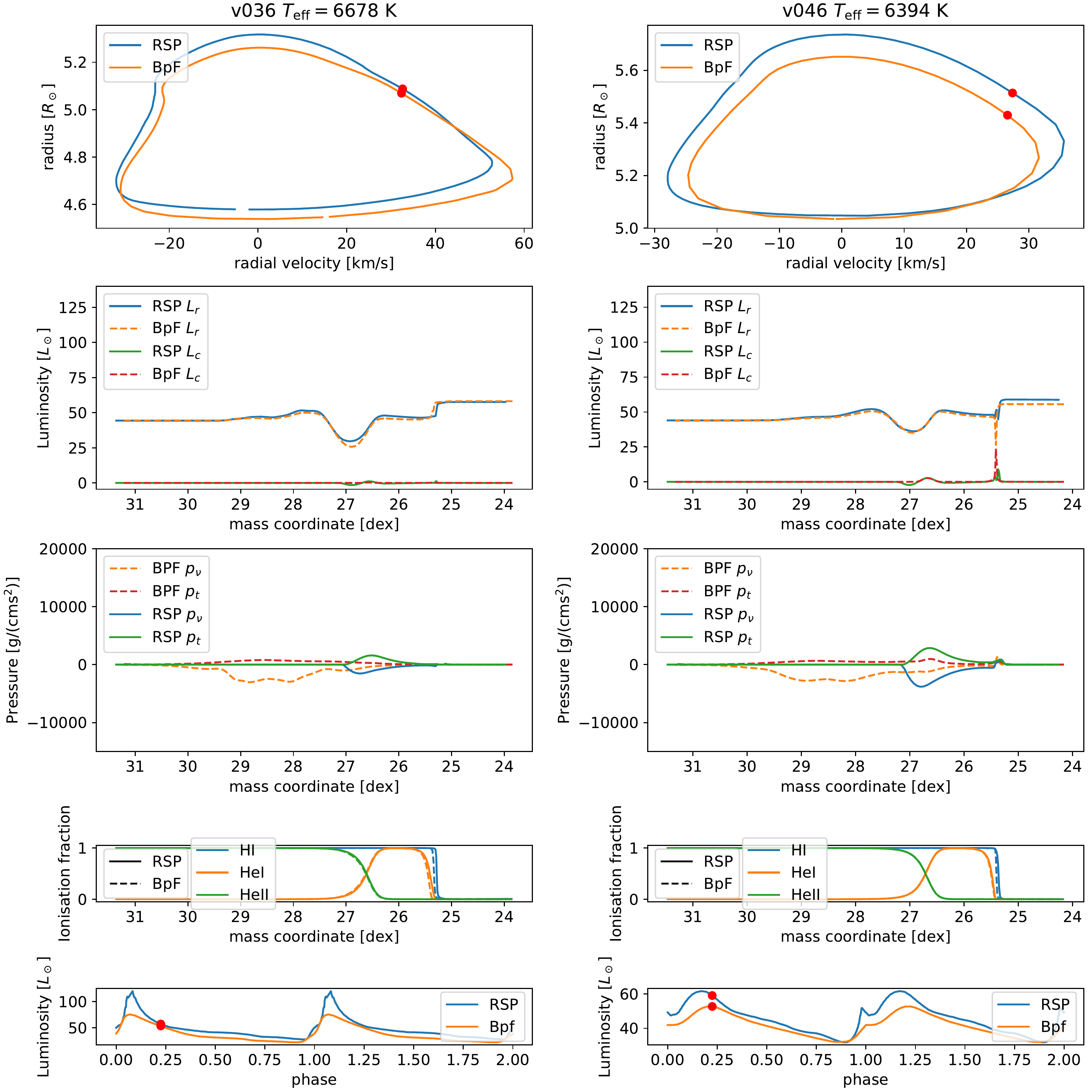}
    \caption{Snapshot from the video attached to this paper, at the decelerating phase. Notations and panels are the same as in Fig.~\ref{fig:snapshot0}. Here we can see that the overshooted $p_t$ and $p_\nu$ are significant compared to the values in the ionization regions.}
    \label{fig:snapshot022}
\end{figure*}

The limit cycle is compared by considering the detailed structural variations 
of the internal regions of the nonlinear models. To better understand this, we attached a  full animation as online supplementary material and included two snapshots in this paper at two pulsation phases. The attached animation shows the time evolution of the internal structure of the two models for two pulsation cycles. Fig. \ref{fig:snapshot0} shows one snapshot taken at a minimum radius, and Fig.\ref{fig:snapshot022} shows the other snapshot in the decelerating (between $v_{\rm max}$ and $R_{\rm max}$) phase. The convective parameters were chosen as in Section~\ref{sec:comp-fits}, i.e. according to the star's regression lines (see Table. \ref{tab:regression_coefficients}), and the fitted parameters (Table \ref{tab:parameters}). 

In the top panel of Fig. \ref{fig:snapshot0} we can see that the \rsp\ model has a slightly larger radius throughout the pulsation phase; in the second panel, one can see that for higher temperature stars in the \bpf\ the convective parameters are smaller than in the \rsp\ model. 

In the minimum radius phase, the ionization regions of the  stars are fully convective. (The ionization region can be seen in the second panels from the bottom, where the ionization fraction changes from 0 to 1.) In the second panel, one can see that in the case of the \rsp\ the negative buoyancy forces support a backward (negative) convection at the base of the HeII ionization region. The convective luminosity in this region's lower temperature model (right panel) is equal in both models. One can also see the high luminosity built-up of energy reserve in the ionization region. This is created in the contraction phase when   a large pressure gradient needs to decelerate the in-falling material. The ionization fronts also move due to interaction with the radiation field in this phase. Thus these parts of the star have more significant temperatures and pressure, accelerating the outer shells outwards rapidly up to the maximum velocity when the opacity can drop and the energy is finally released.

We can see in the panel second from the bottom that the models' ionization regions have almost identical shapes, although the pressure terms (middle panels) connected to the convection are different. The  turbulent and eddy viscous pressure ($p_t$ and $p_\nu$) profiles, caused by the neglection of the negative $\bar{Y}$, are diffused out from the convective region in the \bpf . Because of this, they have different shapes and lower peaks compared to the \rsp .

This overshooting diffuse $p_\nu$ and $p_t$ starts to be very problematic in the decelerating phase (i.e., between maximum velocity and maximum radius), which can be seen in Fig.~\ref{fig:snapshot022}. The deceleration is slow, so the assumptions of the quasi-static atmosphere are best fulfilled here \citep[for the use of atmospheric models with pulsation codes, see][]{Bono1994}. 

One powerful feature is the presence of turbulent and eddy viscosity terms in regions outside the convective regions. This is the outcome of the $e_t$ overshooting mentioned in the previous section. This is relatively small when the convection is strong but not damped alongside the convection in the slowing down phases. Between the phase of the disappearing and reappearing convection fluxes, these terms are very prominent, having an almost constant work on the star.
The difference between the $L_c$ and $L_r$ profiles disappeared in the \rsp , suggesting that the static envelope inconsistency has been resolved for the limit cycle. 

\section{Discussion}
\label{sec:discussion}

Based on the previously described method, we fitted the numerical values of $\alpha$ parameters contained in the two 1D numerical codes describing the time-dependent turbulent convection by slightly different parameterized physical models to the measured radial velocities of several stable RRab stars of M3 globular cluster. This procedure has got to values that we recommend to model RRab stars with the two hydrocodes.

\subsection{Recommended $\alpha$ parameters}
\label{sec:recommendend_alpha}

The calibration showed that there is a connection between  the eddy viscosity scale, the turbulent dissipation efficiency, and the effective temperature:  $\bar{\alpha}_\nu$-$\bar{\alpha}_d$-$T_\textrm{eff}$. As the phenomenon is strongly connected to the temperature gradients, and thus the temperature itself, this is not surprising.  Moreover, this  strengthens the findings of \cite{Bono2002,DiCriscienzo2004,DiCriscienzo2004a}, where they stated that closer to the red edge of the instability strip one would need a larger mixing length parameter (increasing from 1.5 to 1.8--2, although their parametrization is different from ours), meaning larger convection efficiency.  However, it means a degeneracy between $\bar{\alpha}_\nu$-$\bar{\alpha}_d$. 
This reduces the number of free parameters and makes the fit of one of them impossible for our sample. Hence to suggest a value for $\bar{\alpha}_d$ we argue that one should use a value 
that is small enough to decrease the dependence of the $T_{\rm eff}$. Hence in the case of the \rsp\ , we recommend using the standard Kuhfuss value $\bar{\alpha}_d=8/3\sqrt{2/3}\approx 2.17$\footnote{This is the original value derived by \cite{Kuhfuss1986}, and is also present in all \rsp\ prescriptions of \cite{Paxton2019}.}. In the case of the \bpf, the regression is only significant $\bar{\alpha}_d \gtrsim 9$, so we choose $\bar{\alpha}_d = 10.6$, which is the median of the best fitting region ($\bar{\alpha}_d \in [9.6,11.6]$ described in Sec.~\ref{sec:bpf-results}). In Table~\ref{tab:best_parameters},we present the $\bar{\alpha}_\nu$ suggestions based on the expressions (\ref{eq:a_d_a_nu_t_bpf}) and (\ref{eq_a_d_a_nu_t_rsp}). 

As a result of the parameter calibration of \bpf\ we found two parameter sets as local minima called $\mathfrak{A}$ and $\mathfrak{B}$ (see Table~\ref{tab:a_c_a_s_sets}). While parameter set $\mathfrak{A}$ has slightly better $\chi^2$-s, choosing between the two is impossible through $\chi^2$ as they technically describe two  different physical models. In the case of set   $\mathfrak{A}$, the source of the turbulent energy is solely the effects of the Reynold stresses (namely $p_t$ and $p_\nu$). This scenario is implausible since the turbulent energy itself results from fulfilling the Schwarzshcild-criterion, which is embedded into the coupling term $\mathcal{C}$. In addition, some derivation completely misses $\bar{\alpha}_s$ taking it equal to $\bar{\alpha}_c$ \citep{Kuhfuss1986,lengyel}. This reasoning is the cause that we drop parameter set  $\mathfrak{A}$ from our suggested parameters in Table~\ref{tab:best_parameters}.

As  the turbulent pressure parameter, $\bar{\alpha}_p$ has little effect on the radial-velocity curves, we suggest using $\bar{\alpha}_p=2/3$ which is the previous suggestion of \citet{bpf-beat2002}. To get more precise values for $\bar{\alpha}_t$, $\bar{\alpha}_p$, and perhaps $\bar{\alpha}_d$, one possibility is the calibration through modeling mode selection in RRab stars. A calibration for RRc stars is also in preparation, which can influence the results of such calibration.

One can argue that fitting to light curves can give better results, as light curves are more sensitive to convective parameters. While this is true, we point out that light curve fitting is more error-prone since the models do not describe the atmosphere. Thus, we can rely only on bolometric corrections; meanwhile, the  luminosities are also sensitive to the distance modulus. Despite this, we can see in Fig.~\ref{fig:best_fit_curves} that in some cases, the fitting of the RV curve gives a reasonably good fit for some parts of the light curves (v009, v018, v036, v083 in the \bpf\ case.) Otherwise, fitting to the light curve would give unrealistically low RV amplitudes when we are close to the blue edge of the IS.  While this problem can be mitigated by choosing smaller p-factors, this option is less usable for RR Lyrae stars (as their p-factor do not vary that much, see Sec. \ref{sec:observations}) and even in the case of Cepheids could lead to unrealistic p-factors \citep{Marconi2013b,Marconi2017}. Features like the too-strong bumps in the \rsp\ models are unavoidable  even with carefully chosen parameters (while maintaining a fit for the amplitudes). This is similar to the results of \cite{Bono2000}, where they compared model lightcurves of U Comae calculated with and without negative buoyancy, and they also found that the bump is too strong in the latter case.

We also note that if we fit different parts of the RV (or the light) curve, we can end up with different values for the fitting, which can be interpreted as a phase dependence described by \citet{Mundprecht2015}. However, in the current form of the models, phase-dependence is misleading since we always have a full-amplitude model evolved using a fixed parameter set. In the case of the light curves, the different choices of the parameters cause a shift in the observable features like the bumps; this way, one can eliminate the problem of the bumps by simply choosing parameters that shift it away. At the same time, one can reproduce the bump at the desired location simply by changing the amplitude and hence the slope of the curve.

It is worth pointing out why previous prescriptions of the $\bar{\alpha}$-s gave acceptable results. Those former prescriptions were based on trial-by-error searches \citep[see][]{Yecko1998,bpf-beat2002} or arguments to comfort standard Mixing Length Theory \citep{Kuhfuss1986,lengyel}. We can see in Fig.~\ref{fig:compared_regressions} that the \bpf\ parameters (regarding the $\bar{\alpha}_\nu$) are in the middle of the region, which is covered by the regression lines, while the $\bar{\alpha}_s$ and $\bar{\alpha}_c$ parameters are also close to the fitted values.
On the other hand, the possible $\bar{\alpha}_\nu$ parameters cover a larger region of the parameter space in the case of the \rsp, including the previous prescriptions given by \citet{Paxton2019}.

\subsection{The differences between the numerical model structures}
\label{sec:discuss-diff}

As we have described in Section \ref{sec:comp-fits},  moving towards lower ($T_{\rm eff} <6500$ K) effective temperatures, the \bpf\  synthetic light curves tend to be less precise, while the goodness of the velocity curves improves at the minima for both codes. This tendency hints that the energy transfer model is incomplete but slightly detached from the dynamics. 

In Section \ref{sec:comp-static}, we saw that the turbulent energy profile has very sharp edges at the boundaries of the convective zones in the \rsp, but it is diffuse in the case of the \bpf. \citet{lengyel} interpreted this latter effect as an artificial overshooting arising from the negligence of the negative buoyancy forces in Eq.~(\ref{eq:Y_bar}). Since the damping, $D \propto e_t^{3/2}$ and the source $S \propto e_t^{1/2}$, the former decreases faster, and thus $e_t$ does not disappear entirely. If one allows $\bar{Y} < 0$, one will get $S<0$, i.e. a negative source, which then continues to deplete $e_t$. After a few pulsation cycles, this creates a non-negligible ($e_t\sim 10^9$ erg) turbulent energy reserve outside the convective region \citep[for more details of this phenomenon, see][section 5.3 and Fig 12.]{lengyel}

Thus, $p_t$ and $p_\nu$ from the remaining $e_t$ will represent extra damping since it will work even when convection is inefficient (technically $\bar{Y} \sim 0$ in the deceleration phase). In some cases, this results in a better light curve in amplitude and morphology, suggesting that a similar effect is needed to describe convection, even if this straightforward solution is physically inconsistent. 

On the other hand, convective overshooting does not occur at all in the \rsp\ code  because the negative source term is relatively large compared to the turbulent flux. We see this abrupt truncation in $e_t$ in the Fig.~\ref{fig:v036_static}. This model physically implies that the convective eddy velocity slows down from a few km/s to 0, as if it were to "slam" into a wall which scenario is also unlikely. This is contradicted by the fact that \cite{Guo2021} has shown in their numerical stellar envelope calculations for static main-sequence A stars that there is an overshooting layer around the ionization zone. However, the temperature of RR Lyrae stars is slightly lower in comparison, so there must be more potent convective effects, which are even stirred up by the pulsation.

Nevertheless, both numerical codes can track the pulsation dynamics, while the description of the energy transfer seems incomplete. The appropriate values of $\alpha_d$-$\alpha_\nu$ can achieve this as $p_\nu$ depends directly on the velocity, and its work typically converts kinetic energy into turbulent energy. Contrary, the damping term $D$ works against this, converting turbulent energy into heat, allowing a proper balance between kinetic and turbulent energy. On the other hand, the light curves strongly depend on the surface temperature, a function of internal energy. Since the transfer between the internal and kinetic energy components happens only through particular interactions, we do not have a separate parameter for the feedback, so an incomplete equilibrium determines the outgoing radiation.

As far as \bpf\ seemingly causes a non-physical overshoot, the \rsp\ underestimates the effect of the overshoot, so it may be worthwhile in the future to adjust 1D pulsation codes that can handle this phenomenon properly \citep[see, e.g.][]{Kupka2022}. In addition to the theoretical works required for a better understanding, another line of development might expect progress in numerical studies based on multidimensional models. \cite{Mundprecht2015} found that $\bar{\alpha}_c$ is not constant in time and takes a different value in convectively stable regions and overgrowth layers. This, together with our $\bar{\alpha}_\nu$ temperature dependence, implies that a description of the $\alpha$'s as a function of other physical parameters may be necessary.

Our results also mean that while modeling of light curves is also possible with these codes
 \citep{Bono2000,Bono2002, DiFabrizio2002,KellerWood2002,Marconi2005,Marconi2007} , the light curve-calibrated models will have too low radial-velocity amplitudes for high temperature stars, unless varying the projection factor which method can also lead to unrealistically low ($p\sim 1$) values \citep{Marconi2013b,Marconi2017} . Meanwhile, on the ascending phase of the light curve, the strong acceleration and rapid changes in the stellar structure suggest that the assumptions of a quasi-static atmosphere are invalid in these phase intervals.

\section{Summary}
\label{sec:summary}

We compared and discussed two available 1D pulsation codes in this study. One of them \citep[Budapest-Florida code or \bpf][]{bpf-beat2002}  was the first one, among the codes including time-dependent turbulent convection, to produce double-mode pulsation in the types of classical pulsators. The roots of the other code \citep[MESA-RSP or \rsp ][]{lengyel} are the same, but it applies a bit different description of turbulent convection in the ionization zones, which drives and damp the pulsation of these variable stars. The models contain 8 parameters which give large freedom in their application as a black box, and only some suggestions were given regarding their values. The \rsp\ code was made publicly available \citep{Paxton2019} and is used by several authors to model radially pulsating stars using one of the suggested parameter sets. 

However, these codes are based on simplified theoretical models of turbulent convection, a 3-dimensional and stochastic process in a medium where radiation is transported from the stellar core to the surface and then escapes from there, finally hitting our detectors. Our numerical experiences  showed us that the codes could reproduce the main features of the observations, but they failed in some details. Hence, we performed a detailed comparison of model computations for several selected stable RR Lyrae stars of the globular cluster M3, where we have precise photometric and radial velocity measurements \citep{Jurcsik2017}. The photometric data in different spectral bands cannot be compared directly to the model computations because the latter gives only the integrated bolometric values. So, we turned to the radial velocities, which are precise and measured in absolute units. We can fit them directly to model computations which provide the variation of the radius of the models. Of course, the projection factor connects the two data sets and contains some uncertainty, but this scaling factor is well-controlled. 

In this work, we confronted the computed and measured radial velocities of these stars and searched the best fitting of the otherwise undefined $\alpha$ values contained in the description of turbulent convection. We have got their best-fitting values and some correlation between them and their dependence on the effective temperature of the studied stars. Although this specification narrowed down the deviation in the details of radial velocities, the codes cannot provide the observed luminosity variations parallel with comparable accuracy. This discrepancy means that both codes have some missing physical processes in the energy transport toward the stellar surface and the working of pulsating stars. 

However, because the fixation to radial velocities is independent of these mechanisms, the $\alpha$ values may help  apply these codes to model observations. The recommended values are summarised in Table~\ref{tab:best_parameters}. To make this parameter recommendation distinct from the other prescriptions in the literature, we call this the RRab set. 

This work is part of a more extended effort to improve modeling the classical radially pulsating stars through calibrations and comparison with multi-dimensional and extended modeling results.

\begin{table}
    \centering
    \caption{Recommended parameter sets for RRab stars. The given errors show the uncertainty of the calibration. The uncertainty of $\bar{\alpha}_\nu$ is described in detail in Section \ref{sec:errors}. }
    \begin{tabular}{c|c|c}
        parameter & \rsp & \bpf \\
        \hline
        $\Bar{\alpha}_\Lambda$ & 1.5 & 1.5\\
        $\bar{\alpha}_\nu$ &  $-8.65\log T_\textrm{eff} +33.38$ & $-6.77 \log T_\textrm{eff} + 26.09$\\
        $\bar{\alpha}_t$ &  $0.24 \pm 0.03$ & $0.2733$\\
        $\bar{\alpha}_p$ &  $2/3$ & $2/3$\\
        $\bar{\alpha}_d$ &  $8/3\sqrt{2/3}$ & $10.6$\\
        $\bar{\alpha}_s$ & $0.31 \pm 0.07$ & $0.22 \pm 0.08$ \\
        $\bar{\alpha}_c$ &  $0.27\pm 0.03$ & $0.17\pm 0.02$\\
        $\bar{\alpha}_r$ & $0$ & $0$\\
    \end{tabular}
    
    \label{tab:best_parameters}
\end{table}

\section*{Acknowledgements}

We are grateful to the anonymous referee for the careful and thorough reading of our manuscript, and whose suggestions greatly improved the quality of this paper.

This project has been supported by the Lend\"ulet Program  of the Hungarian Academy of Sciences, project No. LP2018-7/2022, the `SeismoLab' KKP-137523 \'Elvonal, OTKA projects K-129249 and NN-129075, as well as the MW-Gaia COST Action (CA18104).

On behalf of Project 'Hydrodynamical modeling of classical pulsating variables with SPHERLS' we are grateful for the usage of ELKH Cloud \citep[see][\url{https://science-cloud.hu/}]{H_der_2022} which helped us achieve the results published in this paper.

\section*{Data Availability}

The observations used in this work are publicly available from the online material of \citet{Jurcsik2017}, the \rsp\ numerical code is also available publicly as part of the \texttt{MESA} software \citep{Paxton2019}.

\bibliographystyle{mnras}
\bibliography{master}

\begin{thebibliography}{}
\makeatletter
\relax
\def\mn@urlcharsother{\let\do\@makeother \do\$\do\&\do\#\do\^\do\_\do\%\do\~}
\def\mn@doi{\begingroup\mn@urlcharsother \@ifnextchar [ {\mn@doi@}
  {\mn@doi@[]}}
\def\mn@doi@[#1]#2{\def\@tempa{#1}\ifx\@tempa\@empty \href
  {http://dx.doi.org/#2} {doi:#2}\else \href {http://dx.doi.org/#2} {#1}\fi
  \endgroup}
\def\mn@eprint#1#2{\mn@eprint@#1:#2::\@nil}
\def\mn@eprint@arXiv#1{\href {http://arxiv.org/abs/#1} {{\tt arXiv:#1}}}
\def\mn@eprint@dblp#1{\href {http://dblp.uni-trier.de/rec/bibtex/#1.xml}
  {dblp:#1}}
\def\mn@eprint@#1:#2:#3:#4\@nil{\def\@tempa {#1}\def\@tempb {#2}\def\@tempc
  {#3}\ifx \@tempc \@empty \let \@tempc \@tempb \let \@tempb \@tempa \fi \ifx
  \@tempb \@empty \def\@tempb {arXiv}\fi \@ifundefined
  {mn@eprint@\@tempb}{\@tempb:\@tempc}{\expandafter \expandafter \csname
  mn@eprint@\@tempb\endcsname \expandafter{\@tempc}}}

\bibitem[\protect\citeauthoryear{{Anderson}}{{Anderson}}{2018}]{Anderson2018}
{Anderson} R.~I.,  2018, in {Smolec} R.,  {Kinemuchi} K.,   {Anderson} R.~I.,
  eds, ~ Vol. 6, The RR Lyrae 2017 Conference. Revival of the Classical
  Pulsators: from Galactic Structure to Stellar Interior Diagnostics. pp
  193--200 (\mn@eprint {arXiv} {1712.02097})

\bibitem[\protect\citeauthoryear{{Baker}}{{Baker}}{1987}]{Baker1987}
{Baker} N.~H.,  1987, in {Hillebrandt} W.,  {Meyer-Hofmeister} E.,  {Thomas}
  H.~C.,   {Kippenhahn} R.,  eds, Physical Processes in Comets, Stars and
  Active Galaxies. pp 105--124

\bibitem[\protect\citeauthoryear{{Baker} \& {Kippenhahn}}{{Baker} \&
  {Kippenhahn}}{1965}]{Baker-Kippenhahn1965}
{Baker} N.,  {Kippenhahn} R.,  1965, \mn@doi [\apj] {10.1086/148359}, \href
  {https://ui.adsabs.harvard.edu/abs/1965ApJ...142..868B} {142, 868}

\bibitem[\protect\citeauthoryear{{B{\"o}hm-Vitense}}{{B{\"o}hm-Vitense}}{1958}]{mlt}
{B{\"o}hm-Vitense} E.,  1958, \zap, \href
  {https://ui.adsabs.harvard.edu/abs/1958ZA.....46..108B} {46, 108}

\bibitem[\protect\citeauthoryear{{Bono} \& {Stellingwerf}}{{Bono} \&
  {Stellingwerf}}{1994}]{Bono1994}
{Bono} G.,  {Stellingwerf} R.~F.,  1994, \mn@doi [\apjs] {10.1086/192054},
  \href {https://ui.adsabs.harvard.edu/abs/1994ApJS...93..233B} {93, 233}

\bibitem[\protect\citeauthoryear{{Bono}, {Castellani}  \& {Marconi}}{{Bono}
  et~al.}{2000}]{Bono2000}
{Bono} G.,  {Castellani} V.,   {Marconi} M.,  2000, \mn@doi [\apj]
  {10.1086/308263}, \href
  {https://ui.adsabs.harvard.edu/abs/2000ApJ...529..293B} {529, 293}

\bibitem[\protect\citeauthoryear{{Bono}, {Castellani}  \& {Marconi}}{{Bono}
  et~al.}{2002}]{Bono2002}
{Bono} G.,  {Castellani} V.,   {Marconi} M.,  2002, \mn@doi [\apjl]
  {10.1086/339420}, \href
  {https://ui.adsabs.harvard.edu/abs/2002ApJ...565L..83B} {565, L83}

\bibitem[\protect\citeauthoryear{{Buchler}, {Koll{\'a}th}  \&
  {Marom}}{{Buchler} et~al.}{1997}]{tcmodel}
{Buchler} J.~R.,  {Koll{\'a}th} Z.,   {Marom} A.,  1997, \mn@doi [ApSS]
  {10.1023/A:1000564401155}, \href
  {https://ui.adsabs.harvard.edu/abs/1997Ap%26SS.253..139B} {253, 139}

\bibitem[\protect\citeauthoryear{Cacciari, Corwin  \& Carney}{Cacciari
  et~al.}{2005}]{Cacciari2005}
Cacciari C.,  Corwin T.~M.,   Carney B.~W.,  2005, \mn@doi [The Astronomical
  Journal] {10.1086/426325}, 129, 267

\bibitem[\protect\citeauthoryear{{Castor}}{{Castor}}{1968}]{Castor1968}
{Castor} J.~I.,  1968, unpublished preprint

\bibitem[\protect\citeauthoryear{{Castor}}{{Castor}}{1971}]{Castor1971}
{Castor} J.~I.,  1971, \mn@doi [\apj] {10.1086/150945}, \href
  {https://ui.adsabs.harvard.edu/abs/1971ApJ...166..109C} {166, 109}

\bibitem[\protect\citeauthoryear{{Catelan} \& {Smith}}{{Catelan} \&
  {Smith}}{2015}]{Catelan2015_book}
{Catelan} M.,  {Smith} H.~A.,  2015, {Pulsating Stars}.
Wiley-VCH

\bibitem[\protect\citeauthoryear{{Christy}}{{Christy}}{1964}]{Christy1964}
{Christy} R.~F.,  1964, \mn@doi [Reviews of Modern Physics]
  {10.1103/RevModPhys.36.555}, \href
  {https://ui.adsabs.harvard.edu/abs/1964RvMP...36..555C} {36, 555}

\bibitem[\protect\citeauthoryear{{Deupree}}{{Deupree}}{1977a}]{Deupree1977a}
{Deupree} R.~G.,  1977a, \mn@doi [\apj] {10.1086/154958}, \href
  {https://ui.adsabs.harvard.edu/abs/1977ApJ...211..509D} {211, 509}

\bibitem[\protect\citeauthoryear{{Deupree}}{{Deupree}}{1977b}]{Deupree1977b}
{Deupree} R.~G.,  1977b, \mn@doi [\apj] {10.1086/155276}, \href
  {https://ui.adsabs.harvard.edu/abs/1977ApJ...214..502D} {214, 502}

\bibitem[\protect\citeauthoryear{{Deupree}}{{Deupree}}{1977c}]{Deupree1977c}
{Deupree} R.~G.,  1977c, \mn@doi [\apj] {10.1086/155352}, \href
  {https://ui.adsabs.harvard.edu/abs/1977ApJ...215..232D} {215, 232}

\bibitem[\protect\citeauthoryear{{Deupree}}{{Deupree}}{1977d}]{Deupree1977d}
{Deupree} R.~G.,  1977d, \mn@doi [\apj] {10.1086/155396}, \href
  {https://ui.adsabs.harvard.edu/abs/1977ApJ...215..620D} {215, 620}

\bibitem[\protect\citeauthoryear{{Di Criscienzo}, {Marconi}  \& {Caputo}}{{Di
  Criscienzo} et~al.}{2004a}]{DiCriscienzo2004}
{Di Criscienzo} M.,  {Marconi} M.,   {Caputo} F.,  2004a, \memsai, \href
  {https://ui.adsabs.harvard.edu/abs/2004MmSAI..75..190D} {75, 190}

\bibitem[\protect\citeauthoryear{{Di Criscienzo}, {Marconi}  \& {Caputo}}{{Di
  Criscienzo} et~al.}{2004b}]{DiCriscienzo2004a}
{Di Criscienzo} M.,  {Marconi} M.,   {Caputo} F.,  2004b, \mn@doi [\apj]
  {10.1086/422742}, \href
  {https://ui.adsabs.harvard.edu/abs/2004ApJ...612.1092D} {612, 1092}

\bibitem[\protect\citeauthoryear{{Di Criscienzo} et~al.,}{{Di Criscienzo}
  et~al.}{2011}]{DiCriscienzo2011}
{Di Criscienzo} M.,  et~al., 2011, \mn@doi [\aj] {10.1088/0004-6256/141/3/81},
  \href {https://ui.adsabs.harvard.edu/abs/2011AJ....141...81D} {141, 81}

\bibitem[\protect\citeauthoryear{{Di Fabrizio} et~al.,}{{Di Fabrizio}
  et~al.}{2002}]{DiFabrizio2002}
{Di Fabrizio} L.,  et~al., 2002, \mn@doi [\mnras]
  {10.1046/j.1365-8711.2002.05824.x}, \href
  {https://ui.adsabs.harvard.edu/abs/2002MNRAS.336..841D} {336, 841}

\bibitem[\protect\citeauthoryear{{Dorfi} \& {Feuchtinger}}{{Dorfi} \&
  {Feuchtinger}}{1999}]{DorfiFeucthinger1999}
{Dorfi} E.~A.,  {Feuchtinger} M.~U.,  1999, \aap, \href
  {https://ui.adsabs.harvard.edu/abs/1999A&A...348..815D} {348, 815}

\bibitem[\protect\citeauthoryear{{Feuchtinger}}{{Feuchtinger}}{1999}]{Vienna_Feuchtinger}
{Feuchtinger} M.~U.,  1999, \mn@doi [\aaps] {10.1051/aas:1999462}, \href
  {https://ui.adsabs.harvard.edu/abs/1999A&AS..136..217F} {136, 217}

\bibitem[\protect\citeauthoryear{{Freytag}, {Steffen}, {Ludwig},
  {Wedemeyer-B{\"o}hm}, {Schaffenberger}  \& {Steiner}}{{Freytag}
  et~al.}{2012}]{Freytag2012}
{Freytag} B.,  {Steffen} M.,  {Ludwig} H.~G.,  {Wedemeyer-B{\"o}hm} S.,
  {Schaffenberger} W.,   {Steiner} O.,  2012, \mn@doi [Journal of Computational
  Physics] {10.1016/j.jcp.2011.09.026}, \href
  {https://ui.adsabs.harvard.edu/abs/2012JCoPh.231..919F} {231, 919}

\bibitem[\protect\citeauthoryear{{Gehmeyr} \& {Winkler}}{{Gehmeyr} \&
  {Winkler}}{1992}]{GW1992}
{Gehmeyr} M.,  {Winkler} K. H.~A.,  1992, \aap, \href
  {https://ui.adsabs.harvard.edu/abs/1992A&A...253...92G} {253, 92}

\bibitem[\protect\citeauthoryear{{Geroux} \& {Deupree}}{{Geroux} \&
  {Deupree}}{2011}]{SPHERLS1}
{Geroux} C.~M.,  {Deupree} R.~G.,  2011, \mn@doi [ApJ]
  {10.1088/0004-637X/731/1/18}, \href
  {https://ui.adsabs.harvard.edu/abs/2011ApJ...731...18G} {731, 18}

\bibitem[\protect\citeauthoryear{{Geroux} \& {Deupree}}{{Geroux} \&
  {Deupree}}{2014}]{SPHERLSIII}
{Geroux} C.~M.,  {Deupree} R.~G.,  2014, \mn@doi [\apj]
  {10.1088/0004-637X/783/2/107}, \href
  {https://ui.adsabs.harvard.edu/abs/2014ApJ...783..107G} {783, 107}

\bibitem[\protect\citeauthoryear{{Geroux} \& {Deupree}}{{Geroux} \&
  {Deupree}}{2015}]{SPHERLS4}
{Geroux} C.~M.,  {Deupree} R.~G.,  2015, \mn@doi [ApJ]
  {10.1088/0004-637X/800/1/35}, \href
  {https://ui.adsabs.harvard.edu/abs/2015ApJ...800...35} {800, 35}

\bibitem[\protect\citeauthoryear{{Gough}}{{Gough}}{1977}]{Gough1977}
{Gough} D.~O.,  1977, \mn@doi [\apj] {10.1086/155244}, \href
  {https://ui.adsabs.harvard.edu/abs/1977ApJ...214..196G} {214, 196}

\bibitem[\protect\citeauthoryear{{Guo} \& {Li}}{{Guo} \& {Li}}{2021}]{Guo2021}
{Guo} F.,  {Li} Y.,  2021, \mn@doi [\apj] {10.3847/1538-4357/abe1c5}, \href
  {https://ui.adsabs.harvard.edu/abs/2021ApJ...910...34G} {910, 34}

\bibitem[\protect\citeauthoryear{H{\'{e}}der et~al.,}{H{\'{e}}der
  et~al.}{2022}]{H_der_2022}
H{\'{e}}der M.,  et~al., 2022, \mn@doi [Inform{\'{a}}ci{\'{o}}s
  T{\'{a}}rsadalom] {10.22503/inftars.xxii.2022.2.8}, 22, 128

\bibitem[\protect\citeauthoryear{{Jurcsik}}{{Jurcsik}}{2003}]{Jurcsik2003}
{Jurcsik} J.,  2003, \mn@doi [\aap] {10.1051/0004-6361:20030418}, \href
  {https://ui.adsabs.harvard.edu/abs/2003A&A...403..587J} {403, 587}

\bibitem[\protect\citeauthoryear{{Jurcsik} et~al.,}{{Jurcsik}
  et~al.}{2015}]{Jurcsik2015}
{Jurcsik} J.,  et~al., 2015, \mn@doi [\apjs] {10.1088/0067-0049/219/2/25},
  \href {https://ui.adsabs.harvard.edu/abs/2015ApJS..219...25J} {219, 25}

\bibitem[\protect\citeauthoryear{{Jurcsik} et~al.,}{{Jurcsik}
  et~al.}{2017}]{Jurcsik2017}
{Jurcsik} J.,  et~al., 2017, \mn@doi [\mnras] {10.1093/mnras/stx382}, \href
  {https://ui.adsabs.harvard.edu/abs/2017MNRAS.468.1317J} {468, 1317}

\bibitem[\protect\citeauthoryear{{Keller} \& {Wood}}{{Keller} \&
  {Wood}}{2002}]{KellerWood2002}
{Keller} S.~C.,  {Wood} P.~R.,  2002, \mn@doi [\apj] {10.1086/342315}, \href
  {https://ui.adsabs.harvard.edu/abs/2002ApJ...578..144K} {578, 144}

\bibitem[\protect\citeauthoryear{{Keller} \& {Wood}}{{Keller} \&
  {Wood}}{2006}]{KellerWood2006}
{Keller} S.~C.,  {Wood} P.~R.,  2006, \mn@doi [\apj] {10.1086/501115}, \href
  {https://ui.adsabs.harvard.edu/abs/2006ApJ...642..834K} {642, 834}

\bibitem[\protect\citeauthoryear{{Koll{\'a}th}, {Buchler}, {Szab{\'o}}  \&
  {Csubry}}{{Koll{\'a}th} et~al.}{2002}]{bpf-beat2002}
{Koll{\'a}th} Z.,  {Buchler} J.~R.,  {Szab{\'o}} R.,   {Csubry} Z.,  2002,
  \mn@doi [\aap] {10.1051/0004-6361:20020182}, \href
  {https://ui.adsabs.harvard.edu/abs/2002A&A...385..932K} {385, 932}

\bibitem[\protect\citeauthoryear{{Kuhfuss}}{{Kuhfuss}}{1986}]{Kuhfuss1986}
{Kuhfuss} R.,  1986, \aap, \href
  {https://ui.adsabs.harvard.edu/abs/1986A&A...160..116K} {160, 116}

\bibitem[\protect\citeauthoryear{{Kupka} \& {Muthsam}}{{Kupka} \&
  {Muthsam}}{2017}]{kupka}
{Kupka} F.,  {Muthsam} H.~J.,  2017, \mn@doi [Living Reviews in Computational
  Astrophysics] {10.1007/s41115-017-0001-9}, \href
  {https://ui.adsabs.harvard.edu/abs/2017LRCA....3....1K} {3, 1}

\bibitem[\protect\citeauthoryear{{Kupka}, {Ahlborn}  \& {Weiss}}{{Kupka}
  et~al.}{2022}]{Kupka2022}
{Kupka} F.,  {Ahlborn} F.,   {Weiss} A.,  2022, arXiv e-prints, \href
  {https://ui.adsabs.harvard.edu/abs/2022arXiv220712296K} {p. arXiv:2207.12296}

\bibitem[\protect\citeauthoryear{{Lee} \& {Sneden}}{{Lee} \&
  {Sneden}}{2021}]{Lee2021}
{Lee} J.-W.,  {Sneden} C.,  2021, \mn@doi [\apj] {10.3847/1538-4357/abd948},
  \href {https://ui.adsabs.harvard.edu/abs/2021ApJ...909..167L} {909, 167}

\bibitem[\protect\citeauthoryear{{Marconi} \& {Clementini}}{{Marconi} \&
  {Clementini}}{2005}]{Marconi2005}
{Marconi} M.,  {Clementini} G.,  2005, \mn@doi [\aj] {10.1086/429525}, \href
  {https://ui.adsabs.harvard.edu/abs/2005AJ....129.2257M} {129, 2257}

\bibitem[\protect\citeauthoryear{{Marconi} \& {Degl'Innocenti}}{{Marconi} \&
  {Degl'Innocenti}}{2007}]{Marconi2007}
{Marconi} M.,  {Degl'Innocenti} S.,  2007, \mn@doi [\aap]
  {10.1051/0004-6361:20065840}, \href
  {https://ui.adsabs.harvard.edu/abs/2007A&A...474..557M} {474, 557}

\bibitem[\protect\citeauthoryear{{Marconi}, {Molinaro}, {Ripepi}, {Musella}  \&
  {Brocato}}{{Marconi} et~al.}{2013a}]{Marconi2013b}
{Marconi} M.,  {Molinaro} R.,  {Ripepi} V.,  {Musella} I.,   {Brocato} E.,
  2013a, \mn@doi [\mnras] {10.1093/mnras/sts197}, \href
  {https://ui.adsabs.harvard.edu/abs/2013MNRAS.428.2185M} {428, 2185}

\bibitem[\protect\citeauthoryear{{Marconi} et~al.,}{{Marconi}
  et~al.}{2013b}]{Marconi2013a}
{Marconi} M.,  et~al., 2013b, \mn@doi [\apjl] {10.1088/2041-8205/768/1/L6},
  \href {https://ui.adsabs.harvard.edu/abs/2013ApJ...768L...6M} {768, L6}

\bibitem[\protect\citeauthoryear{{Marconi} et~al.,}{{Marconi}
  et~al.}{2015}]{Marconi2015}
{Marconi} M.,  et~al., 2015, \mn@doi [\apj] {10.1088/0004-637X/808/1/50}, \href
  {https://ui.adsabs.harvard.edu/abs/2015ApJ...808...50M} {808, 50}

\bibitem[\protect\citeauthoryear{{Marconi} et~al.,}{{Marconi}
  et~al.}{2017}]{Marconi2017}
{Marconi} M.,  et~al., 2017, \mn@doi [\mnras] {10.1093/mnras/stw3289}, \href
  {https://ui.adsabs.harvard.edu/abs/2017MNRAS.466.3206M} {466, 3206}

\bibitem[\protect\citeauthoryear{{Marconi}, {Bono}, {Pietrinferni}, {Braga},
  {Castellani}  \& {Stellingwerf}}{{Marconi} et~al.}{2018}]{Marconi2018}
{Marconi} M.,  {Bono} G.,  {Pietrinferni} A.,  {Braga} V.~F.,  {Castellani} M.,
    {Stellingwerf} R.~F.,  2018, \mn@doi [\apjl] {10.3847/2041-8213/aada17},
  \href {https://ui.adsabs.harvard.edu/abs/2018ApJ...864L..13M} {864, L13}

\bibitem[\protect\citeauthoryear{{Molinaro} et~al.,}{{Molinaro}
  et~al.}{2012}]{Molinaro2012}
{Molinaro} R.,  et~al., 2012, \mn@doi [\apj] {10.1088/0004-637X/748/1/69},
  \href {https://ui.adsabs.harvard.edu/abs/2012ApJ...748...69M} {748, 69}

\bibitem[\protect\citeauthoryear{{Mundprecht}, {Muthsam}  \&
  {Kupka}}{{Mundprecht} et~al.}{2013}]{Mundprecht2013}
{Mundprecht} E.,  {Muthsam} H.~J.,   {Kupka} F.,  2013, \mn@doi [\mnras]
  {10.1093/mnras/stt1511}, \href
  {https://ui.adsabs.harvard.edu/abs/2013MNRAS.435.3191M} {435, 3191}

\bibitem[\protect\citeauthoryear{{Mundprecht}, {Muthsam}  \&
  {Kupka}}{{Mundprecht} et~al.}{2015}]{Mundprecht2015}
{Mundprecht} E.,  {Muthsam} H.~J.,   {Kupka} F.,  2015, \mn@doi [\mnras]
  {10.1093/mnras/stv434}, \href
  {https://ui.adsabs.harvard.edu/abs/2015MNRAS.449.2539M} {449, 2539}

\bibitem[\protect\citeauthoryear{{Muthsam}, {Kupka}, {L{\"o}w-Baselli},
  {Obertscheider}, {Langer}  \& {Lenz}}{{Muthsam} et~al.}{2010}]{Muthsam2010}
{Muthsam} H.~J.,  {Kupka} F.,  {L{\"o}w-Baselli} B.,  {Obertscheider} C.,
  {Langer} M.,   {Lenz} P.,  2010, \mn@doi [\na]
  {10.1016/j.newast.2009.12.005}, \href
  {https://ui.adsabs.harvard.edu/abs/2010NewA...15..460M} {15, 460}

\bibitem[\protect\citeauthoryear{{Nardetto}, {Fokin}, {Mourard}, {Mathias},
  {Kervella}  \& {Bersier}}{{Nardetto} et~al.}{2004}]{Nardetto2004}
{Nardetto} N.,  {Fokin} A.,  {Mourard} D.,  {Mathias} P.,  {Kervella} P.,
  {Bersier} D.,  2004, \mn@doi [\aap] {10.1051/0004-6361:20041419}, \href
  {https://ui.adsabs.harvard.edu/abs/2004A&A...428..131N} {428, 131}

\bibitem[\protect\citeauthoryear{{Nardetto}, {Gieren}, {Kervella},
  {Fouqu{\'e}}, {Storm}, {Pietrzynski}, {Mourard}  \& {Queloz}}{{Nardetto}
  et~al.}{2009}]{Nardetto2009}
{Nardetto} N.,  {Gieren} W.,  {Kervella} P.,  {Fouqu{\'e}} P.,  {Storm} J.,
  {Pietrzynski} G.,  {Mourard} D.,   {Queloz} D.,  2009, \mn@doi [\aap]
  {10.1051/0004-6361/200912333}, \href
  {https://ui.adsabs.harvard.edu/abs/2009A&A...502..951N} {502, 951}

\bibitem[\protect\citeauthoryear{{Natale}, {Marconi}  \& {Bono}}{{Natale}
  et~al.}{2008}]{Natale2008}
{Natale} G.,  {Marconi} M.,   {Bono} G.,  2008, \mn@doi [\apjl]
  {10.1086/526518}, \href
  {https://ui.adsabs.harvard.edu/abs/2008ApJ...674L..93N} {674, L93}

\bibitem[\protect\citeauthoryear{{Nordlund}, {Stein}  \& {Asplund}}{{Nordlund}
  et~al.}{2009}]{Nordlund2009}
{Nordlund} {\r{A}}.,  {Stein} R.~F.,   {Asplund} M.,  2009, \mn@doi [Living
  Reviews in Solar Physics] {10.12942/lrsp-2009-2}, \href
  {https://ui.adsabs.harvard.edu/abs/2009LRSP....6....2N} {6, 2}

\bibitem[\protect\citeauthoryear{Paxton et~al.,}{Paxton
  et~al.}{2019}]{Paxton2019}
Paxton B.,  et~al., 2019, \mn@doi [The Astrophysical Journal Supplement Series]
  {10.3847/1538-4365/ab2241}, 243, 10

\bibitem[\protect\citeauthoryear{{Ragosta} et~al.,}{{Ragosta}
  et~al.}{2019}]{Ragosta2019}
{Ragosta} F.,  et~al., 2019, \mn@doi [\mnras] {10.1093/mnras/stz2881}, \href
  {https://ui.adsabs.harvard.edu/abs/2019MNRAS.490.4975R} {490, 4975}

\bibitem[\protect\citeauthoryear{{Sandstrom}, {Pilachowski}  \&
  {Saha}}{{Sandstrom} et~al.}{2001}]{Hydra2001}
{Sandstrom} K.,  {Pilachowski} C.~A.,   {Saha} A.,  2001, \mn@doi [\aj]
  {10.1086/323926}, \href
  {https://ui.adsabs.harvard.edu/abs/2001AJ....122.3212S} {122, 3212}

\bibitem[\protect\citeauthoryear{{Schlafly} \& {Finkbeiner}}{{Schlafly} \&
  {Finkbeiner}}{2011}]{Schlafly2011}
{Schlafly} E.~F.,  {Finkbeiner} D.~P.,  2011, \mn@doi [\apj]
  {10.1088/0004-637X/737/2/103}, \href
  {https://ui.adsabs.harvard.edu/abs/2011ApJ...737..103S} {737, 103}

\bibitem[\protect\citeauthoryear{{Smolec} \& {Moskalik}}{{Smolec} \&
  {Moskalik}}{2008a}]{lengyel}
{Smolec} R.,  {Moskalik} P.,  2008a, ActAA, \href
  {https://ui.adsabs.harvard.edu/abs/2008AcA....58..193S} {58, 193}

\bibitem[\protect\citeauthoryear{{Smolec} \& {Moskalik}}{{Smolec} \&
  {Moskalik}}{2008b}]{lengyel2}
{Smolec} R.,  {Moskalik} P.,  2008b, \actaa, \href
  {https://ui.adsabs.harvard.edu/abs/2008AcA....58..233S} {58, 233}

\bibitem[\protect\citeauthoryear{{Stellingwerf}}{{Stellingwerf}}{1982a}]{Stellingwerf1982a}
{Stellingwerf} R.~F.,  1982a, \mn@doi [\apj] {10.1086/160425}, \href
  {https://ui.adsabs.harvard.edu/abs/1982ApJ...262..330S} {262, 330}

\bibitem[\protect\citeauthoryear{{Stellingwerf}}{{Stellingwerf}}{1982b}]{Stellingwerf1982b}
{Stellingwerf} R.~F.,  1982b, \mn@doi [\apj] {10.1086/160426}, \href
  {https://ui.adsabs.harvard.edu/abs/1982ApJ...262..339S} {262, 339}

\bibitem[\protect\citeauthoryear{{Szab{\'o}}, {Koll{\'a}th}  \&
  {Buchler}}{{Szab{\'o}} et~al.}{2004}]{bpf-drrlyr2004}
{Szab{\'o}} R.,  {Koll{\'a}th} Z.,   {Buchler} J.~R.,  2004, \mn@doi [\aap]
  {10.1051/0004-6361:20035698}, \href
  {https://ui.adsabs.harvard.edu/abs/2004A&A...425..627S} {425, 627}

\bibitem[\protect\citeauthoryear{{Szentgyorgyi} et~al.,}{{Szentgyorgyi}
  et~al.}{2011}]{Szentgyorgy2011}
{Szentgyorgyi} A.,  et~al., 2011, \mn@doi [\pasp] {10.1086/662209}, \href
  {https://ui.adsabs.harvard.edu/abs/2011PASP..123.1188S} {123, 1188}

\bibitem[\protect\citeauthoryear{Torres}{Torres}{2010}]{Torres2010}
Torres G.,  2010, \mn@doi [The Astronomical Journal]
  {10.1088/0004-6256/140/5/1158}, 140, 1158

\bibitem[\protect\citeauthoryear{{Trahin}, {Kervella}, {Gallenne}, {M{\'e}rand}
   \& {Borgniet}}{{Trahin} et~al.}{2018}]{2018pas6.conf..213T}
{Trahin} B.,  {Kervella} P.,  {Gallenne} A.,  {M{\'e}rand} A.,   {Borgniet} S.,
   2018, in {Smolec} R.,  {Kinemuchi} K.,   {Anderson} R.~I.,  eds, ~ Vol. 6,
  The RR Lyrae 2017 Conference. Revival of the Classical Pulsators: from
  Galactic Structure to Stellar Interior Diagnostics. pp 213--216

\bibitem[\protect\citeauthoryear{{Unno}}{{Unno}}{1967}]{Unno1967}
{Unno} W.,  1967, \pasj, \href
  {https://ui.adsabs.harvard.edu/abs/1967PASJ...19..140U} {19, 140}

\bibitem[\protect\citeauthoryear{{Vasilyev}, {Ludwig}, {Freytag}, {Lemasle}  \&
  {Marconi}}{{Vasilyev} et~al.}{2017}]{Vasilyev2017}
{Vasilyev} V.,  {Ludwig} H.~G.,  {Freytag} B.,  {Lemasle} B.,   {Marconi} M.,
  2017, \mn@doi [\aap] {10.1051/0004-6361/201731422}, \href
  {https://ui.adsabs.harvard.edu/abs/2017A&A...606A.140V} {606, A140}

\bibitem[\protect\citeauthoryear{{Vasilyev}, {Ludwig}, {Freytag}, {Lemasle}  \&
  {Marconi}}{{Vasilyev} et~al.}{2018}]{Vasilyev2018}
{Vasilyev} V.,  {Ludwig} H.~G.,  {Freytag} B.,  {Lemasle} B.,   {Marconi} M.,
  2018, \mn@doi [\aap] {10.1051/0004-6361/201732201}, \href
  {https://ui.adsabs.harvard.edu/abs/2018A&A...611A..19V} {611, A19}

\bibitem[\protect\citeauthoryear{{Wood}, {S. Arnold}  \& {Sebo}}{{Wood}
  et~al.}{1997}]{Wood1997}
{Wood} P.~R.,  {S. Arnold} A.,   {Sebo} K.~M.,  1997, \mn@doi [\apjl]
  {10.1086/310798}, \href
  {https://ui.adsabs.harvard.edu/abs/1997ApJ...485L..25W} {485, L25}

\bibitem[\protect\citeauthoryear{{Wuchterl} \& {Feuchtinger}}{{Wuchterl} \&
  {Feuchtinger}}{1998}]{Wuchterl1998}
{Wuchterl} G.,  {Feuchtinger} M.~U.,  1998, \aap, \href
  {https://ui.adsabs.harvard.edu/abs/1998A&A...340..419W} {340, 419}

\bibitem[\protect\citeauthoryear{{Xiong}}{{Xiong}}{1989}]{Xiong1989}
{Xiong} D.-R.,  1989, \aap, \href
  {https://ui.adsabs.harvard.edu/abs/1989A&A...209..126X} {209, 126}

\bibitem[\protect\citeauthoryear{{Yecko}, {Kollath}  \& {Buchler}}{{Yecko}
  et~al.}{1998}]{Yecko1998}
{Yecko} P.~A.,  {Kollath} Z.,   {Buchler} J.~R.,  1998, AAP, \href
  {https://ui.adsabs.harvard.edu/abs/1998A&A...336..553Y} {336, 553}

\makeatother
\end{thebibliography}

\appendix

\section{Unified equations}

To allow a comparison of the two codes we transform their governing equations into a common form. The original equations \citep{bpf-beat2002,lengyel}:

The equation of motion:
\begin{align}
\label{eq:bpf-momentum}
&\textrm{BpF:} & \frac{du}{dt} = - \frac{1}{\rho} \frac{\partial}{\partial r}(p + p_t +p_\nu) - \frac{G M_r}{r^2} \\
\label{eq:rsp-momentum}
&\textrm{RSP:} &\frac{du}{dt} = - \frac{1}{\rho} \frac{\partial}{\partial r}(p +p_t) + U_q - \frac{G M_r}{r^2}
\end{align}
The energy equations:
\begin{align}
    & \textrm{BpF:} & \frac{de}{dt} + p\frac{dv}{dt} = - \frac{1}{\rho r^2}\frac{\partial}{\partial r} (r^2 (F_c + F_r)) + \mathcal{C^\prime }\\
    \label{eq:bpf-t_energy}
    & & \frac{de_t}{dt} + (p_t +p_\nu)\frac{dv}{dt} = - \frac{1}{\rho r^2} \frac{\partial}{\partial r} (r^2 F_t) - \mathcal{C^\prime } \\
    & \textrm{RSP:} & \frac{de}{dt} + p\frac{dv}{dt} = - \frac{1}{\rho r^2}\frac{\partial}{\partial r} (r^2 (F_c + F_r)) - \mathcal{C} \\
    \label{eq:rsp-t_energy}
    &&\frac{de_t}{dt} + p_t\frac{dv}{dt} = - \frac{1}{\rho r^2} \frac{\partial}{\partial r} (r^2 F_t) + E_q + \mathcal{C}
\end{align}
The notations are the same as in Section~\ref{sec:model}. As the coupling term is described differently in the two formalisms, we use the notation $\mathcal{C}^\prime$ in the case of the \bpf. In the case of the \rsp\ model we have two different parameters here: the viscous momentum transfer $U_q$ and viscous energy transfer $E_q$. Their definitions:

\begin{align}
    &    U_q = \frac{1}{\rho r^3}\frac{\partial}{\partial r}\left[ \frac{4}{3} \alpha_m \Lambda \rho e_t^{1/2} r^3\left(\frac{\partial u}{\partial r} - \frac{u}{r}\right)\right] \\
    &  E_q = \frac{4}{3}\frac{1}{\rho} \alpha_m \rho \Lambda e_t^{1/2} \left(\frac{\partial u}{\partial r} - \frac{u}{r}\right)^2  
\end{align}

As the corresponding terms in the \bpf\ model are $\nabla p_\nu$ and $p_\nu \frac{dv}{dt}$ we have to find the connection between these formulae. First if we look at equation (\ref{eq:p_nu}) and take its gradient we can derive a formulation for $U_q(p_\nu)$:
\begin{equation}
    \label{eq:u_q_con}
    U_q = - \frac{1}{\rho} \frac{\partial p_\nu}{\partial r} - \frac{3p_\nu}{\rho r}
\end{equation}
And then we call the second term in this equation $\mathcal{D}_p$.

 For the second connection formula we need to expand $\frac{dv}{dt}$:
 \begin{equation}
     \frac{d v}{dt} = \frac{d}{dt}\left(\frac{1}{\rho}\right) = -\frac{1}{\rho^2} \frac{d \rho}{dt}
 \end{equation}

And we can use the Stokes-derivative and the continuity equation, so $\frac{d \rho}{dt} = -\nabla(\rho u) + u \nabla\rho = \rho\, \textrm{div}\, u$. With this:

\begin{equation}
\label{eq:dvdt_final}
    \frac{d v}{dt} = \frac{1}{\rho} \frac{1}{r^2}\frac{\partial}{\partial r}(r^2 u) =\frac{1}{\rho}\left(\frac{\partial u}{\partial r} - \frac{u}{r}\right)+3\frac{u}{r\rho}
\end{equation}

Multiplying eq.~\ref{eq:dvdt_final} by $p_\nu$ (eq.~\ref{eq:p_nu}) we get the our second connection formula:
\begin{equation}
    \label{eq:E_q_con}
    p_\nu \frac{dv}{dt} = - E_q + 3p_\nu \frac{u}{\rho r}
\end{equation}

Substituting (\ref{eq:u_q_con}) into (\ref{eq:rsp-momentum}) , and (\ref{eq:E_q_con}) into (\ref{eq:rsp-t_energy}) yields equation (\ref{eq:motion}) and (\ref{eq:turbulent_energy}), with the exception of the coupling term.

The next step is the parametrization of the coupling term. In the \rsp\ the term is used as in the main text (eqs. \ref{eq:coupling},\ref{eq:def-source},\ref{eq:def-damp},\ref{eq:dampr},\ref{eq:d_r_fac}, we note that in the original paper of \citet{lengyel} $D_r$ term was our $\mathcal{D}_r$ as there was no need to take out our eq. \ref{eq:d_r_fac}. from the original description of the radiative damping). The coupling term description somewhat differs in \citet{bpf-beat2002},  where:

\begin{align}
        &\mathcal{C^\prime} = \alpha_d \frac{e_t^{1/2}}{\Lambda}(e_t-S_t)\\
        &S_t = (\alpha_s \alpha_\Lambda)^2 e_t^{1/2} \frac{p}{\rho} \beta T Y f_{\rm pec}
\end{align}

Here the main difference is the 
parametrization of $\alpha_s$, the opposite sign in the energy equations, and the  presence of $f_{\rm pec}$, which is the Péclet-correction for radiative damping:
\begin{align}
    f_{\rm pec} = \frac{1}{1+\alpha_r Pe^{-1}},& &Pe = \frac{D_c}{D_R}\\
    D_c= \Lambda e_t^{1/2},& &D_R=\frac{4acT^3}{3\kappa \rho^2 c_p}
\end{align}

If we expand the Péclet number and multiply it by $\alpha_r$, we end up with eq. \ref{eq:d_r_fac}: $D_r = \alpha_r Pe^{-1}$. If one substitutes \citet{lengyel} $\gamma_r^2$ parameter in place of  $\alpha_r$ and multiply $D_r$ by $\frac{3}{4}\frac{e_t^{3/2}}{\Lambda}$ then one gets their damping term ($\mathcal{D}_r^{\mathtt{ RSP}}$ in our notation). 

In the \bpf\ case this correction by the multiplying factor can be converted into an additive factor by using the identity: $\frac{x}{1+y} = x - \frac{xy}{1+y}$. This way, we get:

\begin{equation}
    \mathcal{C^\prime} = \alpha_d \frac{e_t^{1/2}}{\Lambda}\left(e_t-\frac{S_t}{D_r} + \frac{S_t}{f_{\rm pec}}\frac{D_r}{1+D_r} \right)
\end{equation}

After breaking the parenthesis we get the additive formulation for the coupling term:

\begin{equation}
    C^\prime= \alpha_d \frac{e_t^{3/2}}{\Lambda} - \alpha_d \frac{e_t^{1/2}}{\Lambda} \frac{S_t}{f_{\rm pec}} + \alpha_d \frac{e_t^{1/2}}{\Lambda} \frac{S_t}{f_{\rm pec}} \frac{D_r}{1+D_r} 
\end{equation}

Let $\bar{\alpha_s} = \alpha_d\alpha_s^2$, and we can see that:

\begin{equation}
    S = \alpha_d \frac{e_t^{1/2}}{\Lambda} \frac{S_t}{f_{\rm pec}}
\end{equation}
The derivation of the other two terms is straightforward now, and we can see that:

\begin{equation}
    \mathcal{C}^\prime = D - S + \mathcal{D}_r = - \mathcal{C}
\end{equation}

The \bpf\ code also uses the Péclet correction for the convective flux, here we can use the same identity as before:

\begin{equation}
    F_c^{\tt BpF} = F_c f_{\rm pec} = F_c - F_c \frac{D_r}{1+D_r} = F_c - \mathcal{D}_F
\end{equation}

\bsp
\label{lastpage}
\end{document}